\documentclass[twocolumn]{aastex63}
\usepackage{txfonts}
\usepackage{xspace}
\usepackage{longtable}
\usepackage{xcolor}
\usepackage{tabularx}

\usepackage{fancyvrb}

\newcommand\skynet{{\it{SkyNet}}\xspace}
\newcommand{\isotope}[2]{${}^{#1}${#2}}

\newcommand*{\unitstyle}[1]{\mathrm{#1}}
\newcommand*{\Kelvin}{\unitstyle{K}}
\newcommand*{\K}{\Kelvin}  

\newcommand*{\Msun}{M_{\odot}}

\newcommand{\EE}[2]{\ensuremath{{#1}\times 10^{#2}}}

\newcommand\iso[2]{\ensuremath{^{#2}{\rm #1}}}

\newcommand{\thalfref}[2]{$T_{1/2}=#1$~#2~\citep{Wallet11}}

\newcommand{\triplealpha}{$\alpha(2\alpha,\gamma)$\isotope{12}{C}\ }
\newcommand{\triplealphanosp}{$\alpha(2\alpha,\gamma)$\isotope{12}{C}}
\newcommand{\apreac}{($\alpha$,p)}
\newcommand{\agreac}{($\alpha$,$\gamma$)}
\newcommand{\pgreac}{(p,$\gamma$)}
\newcommand{\ngreac}{(n,$\gamma$)}

\newcommand{\pareac}{(p,$\alpha$)}
\submitjournal{ApJ}

\shorttitle{CCSN Sensitivity Study}
\shortauthors{Hermansen et al.}

\begin{document}

\title{Reaction Rate Sensitivity of the Production of $\gamma$-ray Emitting Isotopes in Core-Collapse Supernova}

\author[0000-0002-1707-0844]{Kirby Hermansen}
\affil{National Superconducting Cyclotron Laboratory, Michigan State University, East Lansing, MI 48824, USA}
\affil{Department of Physics and Astronomy, Michigan State University, East Lansing, MI 48824, USA}
\affil{Joint Institute for Nuclear Astrophysics–Center for the Evolution of the Elements, Michigan State University, East Lansing, MI 48824, USA}

\author[0000-0002-5080-5996]{Sean M. Couch}
\affil{National Superconducting Cyclotron Laboratory, Michigan State University, East Lansing, MI 48824, USA}
\affil{Department of Physics and Astronomy, Michigan State University, East Lansing, MI 48824, USA}
\affil{Joint Institute for Nuclear Astrophysics–Center for the Evolution of the Elements, Michigan State University, East Lansing, MI 48824, USA}
\affil{Department of Computational Mathematics, Science, and Engineering, Michigan State University, East Lansing, MI 48824, USA}

\author{Luke F. Roberts}
\affil{National Superconducting Cyclotron Laboratory, Michigan State University, East Lansing, MI 48824, USA}
\affil{Department of Physics and Astronomy, Michigan State University, East Lansing, MI 48824, USA}
\affil{Joint Institute for Nuclear Astrophysics–Center for the Evolution of the Elements, Michigan State University, East Lansing, MI 48824, USA}

\author{Hendrik Schatz}
\affil{National Superconducting Cyclotron Laboratory, Michigan State University, East Lansing, MI 48824, USA}
\affil{Department of Physics and Astronomy, Michigan State University, East Lansing, MI 48824, USA}
\affil{Joint Institute for Nuclear Astrophysics–Center for the Evolution of the Elements, Michigan State University, East Lansing, MI 48824, USA}

\author[0000-0001-9440-6017]{MacKenzie L. Warren}
\altaffiliation{NSF Astronomy and Astrophysics Postdoctoral Fellow}
\affil{Department of Physics and Astronomy, Michigan State University, East Lansing, MI 48824, USA}
\affil{Joint Institute for Nuclear Astrophysics–Center for the Evolution of the Elements, Michigan State University, East Lansing, MI 48824, USA}
\affil{Department of Physics, North Carolina State University, Raleigh, NC 27695, USA}

\begin{abstract}
Radioactive isotopes produced in core-collapse supernovae (CCSNe) provide useful insights into the underlying processes driving the collapse mechanism and the origins of elemental abundances. Their study generates a confluence of major physics research, including experimental measurements of nuclear reaction rates, astrophysical modeling, and $\gamma$-ray observations.
Here we identify the key nuclear reaction rates to the nucleosynthesis of observable radioactive isotopes in explosive silicon-burning during CCSNe. 
Using the nuclear reaction network calculator \skynet and current REACLIB reaction rates, we evolve temperature-density-time profiles of the innermost $0.45~M_\odot$ ejecta from the core collapse and explosion of a $12~M_\odot$ star. Individually varying 3403 reaction rates by factors of 100, we identify 141 reactions which cause significant differences in the isotopes of interest, namely, \isotope{43}{K}, \isotope{47}{Ca}, \isotope{44,47}{Sc}, \isotope{44}{Ti}, \isotope{48,51}{Cr}, \isotope{48,49}{V}, \isotope{52,53}{Mn}, \isotope{55,59}{Fe}, \isotope{56,57}{Co}, and \isotope{56,57,59}{Ni}. For each of these reactions, we present a novel method to extract the temperature range pertinent to the nucleosynthesis of the relevant isotope; the resulting temperatures lie within the range $T = 0.47$ to $6.15~$GK. Limiting the variations to within $1\sigma$ of STARLIB reaction rate uncertainties further reduces the identified reactions to 48 key rates, which can be used to guide future experimental research. Complete results are presented in tabular form.
\end{abstract}



\section{Motivation} \label{sec:intro}

Core collapse supernovae (CCSNe) are important nucleosynthesis sites contributing to the origin of a broad range of elements \citep{Woosley1995, Rauscher2002, Woosley2002, Nomoto2006, Chieffi2017, Curtis2019}. At the extreme conditions during the explosion, many of the synthesized nuclei are produced as radioactive isotopes. Their ultimate contribution to nucleosynthesis is determined by the first stable isotope encountered along their $\beta$- or electron capture decay chains. Most of these isotopes decay quickly, but a few have half-lives that are longer than the explosion timescale and are ejected into the interstellar medium prior to their decay. These longer-lived radioactive isotopes are of particular interest as their signatures can provide information on isotopic abundances that are difficult to obtain from spectroscopic observations and can therefore serve as unique windows into a broad range of physics questions.  These signatures include characteristic decay times of supernova light curves powered by radioactive decay, observation of characteristic $\gamma$-rays emitted in the nuclear decay by balloon or satellite based $\gamma$-ray observatories, isotopic anomalies in geological samples that incorporate supernova ejecta, or the direct detection of the ejected and subsequently accelerated radioactive isotopes as cosmic rays using space based cosmic ray observatories. Our goal here is to delineate the nuclear reactions that need to be understood to reliably predict the production of long-lived radioactive isotopes from CCSN models. This is essential for the interpretation of the observed signatures in terms of CCSN physics. We focus on isotopes produced in explosive oxygen and silicon burning, which is responsible for the synthesis of a broad range of long-lived radioactive isotopes. Explosive silicon burning is of particular interest as it occurs in the deepest layers of the supernova and can therefore provide insights into mixing and ejection mechanisms, and the delineation between ejecta and fall-back onto the compact remnant,  the so-called ``mass-cut" \citep{Young2006, The2006, Grebenev2012, Grefenstette2014}. Other long-lived isotopes in supernovae produced by neutron capture processes in explosive carbon or helium burning layers, such as $^{41}$Ca or $^{60}$Fe, are not the subject of this study.

Nucleosynthesis during explosive silicon burning is governed by a typically $\alpha$-rich freezeout from a quasi-statistical equilibrium \citep{Woosley1992, Hix1996, Meyer1998}. During quasi-statistical equilibrium groups of nuclei on the nuclear chart form equilibrium clusters, where fast nuclear reactions maintain equilibrium among the included nuclei. The resulting relative isotopic abundances within a cluster are therefore entirely determined by the thermodynamic properties of the nuclei and independent of the rates of the nuclear reactions (though the rates of the reactions determine the extent of the cluster). However, equilibrium clusters are connected by slow nuclear reactions that are critical in determining the overall abundance distribution among the clusters. As the material expands, more clusters form and more bottle-neck reactions emerge. 

Sensitivity studies are needed to identify the relatively few critical bottle-neck reactions that affect the final composition. \citet{The1998} performed a sensitivity study using a simple parametrized $\alpha$-rich freezeout model. They identified a number of critical reactions that affect the synthesis of \isotope{44}{Ti} by varying reactions individually by a factor of 100. A similar model was later used to identify nuclear reactions affecting the production of \isotope{59}{Ni}, \isotope{57}{Co}, \isotope{56}{Co}, and \isotope{55}{Fe} \citep{Jordan2003}. 
\citet{Hoffman1999} compared supernova nucleosynthesis in a full 1D explosion model of a 15~$M_\odot$ and a 20~$M_\odot$ star using two sets of reaction rates and identified the $^{40}$Ca($\alpha$,$\gamma$)$^{44}$Ti reaction rate as critical in determining the $^{44}$Ti yield. 
\citet{Hoffman2010} used a set of parametrized expansion models to investigate the sensitivity of $^{44}$Ti production in explosive silicon burning to the $^{40}$Ca($\alpha$,$\gamma$)$^{44}$Ti and $^{44}$Ti($\alpha$,p)$^{47}$V reaction rates over a range of possible conditions. 
\citet{Tur2010} explored the sensitivity of $^{44}$Ti production to the $^{12}$C($\alpha$,$\gamma$)\isotope{16}{O} reaction in three 1D supernova models based on a 15~$M_\odot$, 20~$M_\odot$, and 25~$M_\odot$ progenitor star. \citet{Magkotsios2010} carried out a full sensitivity study of the synthesis of $^{56}$Ni and $^{44}$Ti in explosive silicon burning using a combination of a broad set of parametrized trajectories that systematically cover the relevant parameter space, and trajectories from three supernova models: 
a one-dimensional CasA inspired model with a 16~$M_\odot$ star, a one-dimensional hypernova model, and a two-dimensional rotating 15~$M_\odot$ star model. They varied rates individually by factors of 100.

Here we present a much more comprehensive sensitivity study that considers all radioactive isotopes of potential interest, and uses trajectories from a self-consistent supernova simulation instead of a parametrized approach. 
The CCSN explosion data we use, while one-dimensional, includes the effects of convection and turbulence, accurate energy-dependent neutrino transport, and approximate general relativity.
In addition, we identify for the first time the temperature range over which the model is sensitive to the reaction rate. This information is critical to guide experiments. 

We begin in Section \ref{sec:background} by laying out the background behind detection of radioactive isotopes from CCSNe. Then, in Section \ref{sec:methods} we describe the details of the CCSN simulation, and our post-processing calculations of nucleosynthesis using this simulation. In Section \ref{sec:results} we briefly summarize the results of the study, and describe several points of interest. Next, we contextualize the results in Section \ref{sec:discussion} by evaluating the significance of reactions relative to the uncertainty of their reaction rates. We also compare to past work, both in terms of the nucleosynthesis of the model and the most important reactions identified. Finally, we summarize the key differences in this work in Section \ref{sec:conclusion} and conclude. 

\section{Background} \label{sec:background}
We now discuss the relevant radioactive isotopes and their signatures in more detail. If the half-life of a radioactive isotope produced in a CCSN is long enough to allow mixing outwards to a column depth where $\gamma$-radiation can escape, the decay $\gamma$-radiation can in principle be observed directly with balloon or satellite based $\gamma$-ray detectors \citep{VINK2005, Diehl2017, Timmes2019}.  Such observations provide important isotopic abundance information. In addition, compared to visible light, UV, or X-rays,  $\gamma$-rays are much less affected by attenuation in the surrounding gas or the interstellar medium. It is therefore much more straight forward to determine the total produced abundance, which can then be compared to CCSN model predictions. Due to the challenging instrument sensitivity requirements, only a small number of isotopes have so far been observed via their decay $\gamma$-radiation in supernova remnants. $\gamma$-rays from the decay of \isotope{56}{Co}  (half-life \thalfref{77.236}{d}) were observed from supernova 1987A 160 days after the explosion by balloon experiments \citep{Cook1988, Mahoney1988, Sandie1988, Teegarden1989} and the Solar Maximum Mission (SMM) satellite \citep{Matz1988} (see also the review by \citet{VINK2005}). These observations occurred prior to the expected $\gamma$-ray transparency of the ejecta, indicating the importance of mixing processes during the explosion. Later, the observation of  \isotope{57}{Co} (\thalfref{271.74}{d})  decay $\gamma$-rays from 1987A with the Compton Gamma Ray Observatory (CGRO) was reported by \citet{Kurfess1992}. More recently \isotope{44}{Ti} (\thalfref{60.0}{y}) was detected in 1987A via  $\gamma$- and hard X-rays by INTEGRAL \citep{Grebenev2012} and NuSTAR \citep{Boggs2015}, with total inferred amounts of \isotope{44}{Ti} of
\EE{(3.1 \pm 0.8)}{-4}~$M_\odot$ and  \EE{(1.5 \pm 0.3)}{-4}~$M_\odot$,
respectively. For 1987A, the produced amount of  \isotope{44}{Ti} can also be inferred from the late time light curve \citep{Jerkstrand2011, Seitenzahl2014}. 

The only other supernova remnant from which nuclear decay $\gamma$-rays have been unambiguously detected is Cas A, where $^{44}$Ti has been detected by CGRO/COMPTEL  \citep{Iyudin1994}, BeppoSAX \citep{Vink2001}, INTEGRAL \citep{Renaud2006}, and NuSTAR \citep{Grefenstette2014}. The most recent analysis from INTEGRAL obtains a \isotope{44}{Ti} mass of \EE{(1.37 \pm 0.19)}{-4}~$M_\odot$ \citep{Siegert2015}, while the result from NuSTAR is \EE{(1.25 \pm 0.3)}{-4}~$M_\odot$ \citep{Grefenstette2014}.
These \isotope{44}{Ti} abundances are at least a factor of 3 higher than standard 1D supernova model predictions, which provides constraints on CCSN physics such as homogeneity and the role of multi-D effects such as bi-polar explosions  \citep{The2006, Wheeler2008, Magkotsios2010, Chieffi2017}. No other sources of \isotope{44}{Ti} besides 1987A and CasA have been identified with certainty, though a few candidates with insufficient significance to be considered detections have been reported, e.g., GROJ0852- 4642 (Vela Junior) and G1.9+0.3 (see review by \citet{Diehl2016}). \citet{The2006} argued that the paucity of detectable  \isotope{44}{Ti} in the Galaxy is in conflict with simple assumptions about CCSN thus providing additional constraints on CCSN rates, star formation, and explosion physics. \added{In contrast, \citet{Dufour2013} found the number of supernova remnants with detectable \isotope{44}{Ti} to be consistent with current models, but posited that next generation $\gamma$-ray telescopes \citep{Timmes2019} can be expected to identify between 8 and 21 supernova remnants based on their \isotope{44}{Ti} decay flux. }

\replaced{In principle, $\gamma$-rays from a large number of longer-lived radioactive isotopes could be detectable in the future with a next generation $\gamma$-ray telescope.}{Next generation $\gamma$-ray telescopes are also expected to identify a larger number of longer-lived radioactive isotopes in addition to \isotope{44}{Ti}}. \citet{Timmes2019} estimate that with such an instrument \isotope{48}{Cr}, \isotope{48}{V}, \isotope{52}{Mn}, \isotope{56,57}{Co}, \isotope{56,57}{Ni} may be detectable out to a distance of 1 Mpc, and 
\isotope{43}{K}, \isotope{44}{Ti}, \isotope{44}{Sc}, \isotope{47}{Sc}, \isotope{47}{Ca}, \isotope{51}{Cr}, \isotope{59}{Fe} out to a distance of 50 kpc. Many of these isotopes have half-lives of just hours to days, thus requiring rapid mixing into outer layers where $\gamma$-rays can escape. 

In addition to direct detection of $\gamma$-radiation, signatures of radioactive isotopes from CCSN can be found in geological samples. The analysis of the composition of primitive meteorites provides information about the presence of radioactive isotopes in the early solar system. The main isotope of interest in the context of explosive silicon burning in core collapse supernovae is \isotope{53}{Mn} (\thalfref{3.7}{My}). Analysis of isotopic anomalies created by the presence of the \isotope{53}{Cr} decay daughter in meteorites has provided a fairly accurate value for the early solar system abundance of the \isotope{53}{Mn}/\isotope{55}{Mn} ratio of \EE{6.54 \pm 0.44}{-6} \citep{TISSOT2017}. CCSNe are considered the dominant source, though Type Ia supernovae may also play a role \citep{Wasserburg2006, Cote2019}. 

Isotopes with half-lives in excess of 0.1 My such as \isotope{53}{Mn} accumulate in the interstellar medium, and their early solar system abundance provides a data point for this accumulation at the time and location of solar system formation. This provides unique constraints on chemical evolution probing the more recent galactic chemical history, as opposed to stable isotopes that provide an integrated sample over the age of the Galaxy \citep{Cote2019}. The early solar system abundance of \isotope{53}{Mn} also  probes the circumstances and timescales of solar system formation \citep{Lugaro2018}, in particular the hypothesis 
of late time injection by a supernova that potentially triggered the formation of the solar system \citep{Meyer2000, Wasserburg2006}. With this hypothesis, and using standard spherical CCSN models, there is an overproduction issue of about three orders of magnitude for \isotope{53}{Mn} and \isotope{60}{Fe}
\citep{Meyer2000,Wasserburg2006, Lugaro2018, Banerjee2016}. This has been used to place constraints on the nature of the responsible supernova, e.g. on the layers ejected \citep{Meyer2000}, the nature of fallback \citep{Takigawa2008}, the mass of the progenitor \citep{Banerjee2016}, or the supernova explosion mechanism \citep{Sawada2019}. This underlines the importance of understanding the nuclear production processes. 

Isotopic signatures  of shorter-lived isotopes in supernova ejecta can also be incorporated in geological samples via pre-solar grains. These grains form as SiC dust in the supernova explosion and are then transported through space and incorporated in the solar system, where they can be found in primitive meteorites. 
Indeed, enhanced $^{44}$Ca from the decay of $^{44}$Ti has been found in SiC X-grains thought to originate from CCSNe \citep{Hoppe1996, Nittler1996,  Clayton2011}, see also the recent review by \citet{Nittler2016}. This provides not only constraints on the supernova, but also on the grain formation process.

CCSN signatures of long-lived radioactive isotopes that decay by electron capture can in principle also be identified in the composition of cosmic rays above Earth's atmosphere. After acceleration, the radioactive nuclei are fully stripped of electrons, which prevents electron capture decay and leaves the typically very weak $\beta^+$ decay branches as the only option for decay. As a result the nuclei become sufficiently stable to propagate through the interstellar medium and be detected above Earth's atmosphere \citep{DuVernois1997, Wiedenbeck1999, Neronov2016, Benyamin2018}. The cosmic ray source composition of radioactive isotopes inferred from such observations can therefore serve as a chronometer of the acceleration process. However, the observed composition has to be corrected for secondary production during propagation. Isotopes of interest are \isotope{44}{Ti}, \isotope{49}{V}, \isotope{51}{Cr}, \isotope{55}{Fe}, \isotope{57}{Co}, and \isotope{59}{Ni} \citep{Benyamin2018} as well as \isotope{53}{Mn} \citep{DuVernois1997}. The upper limit on the \isotope{59}{Ni} source abundance obtained from observations with the CRIS instrument on board the Advanced Compton Explorer Spacecraft has placed constraints on cosmic ray acceleration models \citep{ISRAEL2005}. However, \citet{Neronov2016} recently pointed out the critical importance of understanding the production of \isotope{59}{Ni} in explosive silicon burning for discriminating between fast and delayed cosmic ray acceleration models. A signature of \isotope{44}{Ti} in cosmic rays has been reported from CRIS \citep{Scott2005} and has been used to place constraints on \isotope{44}{Ti} synthesis in CCSN \citep{Benyamin2018}. Finally, source limits on \isotope{53}{Mn} have been obtained from data of the Ulysses High Energy Telescope (HET) but uncertainties were too large to draw conclusions \citep{DuVernois1997}. \added{In addition to the detection of the above isotopes by cosmic rays, \citet{Leising2001} estimated the X-rays produced following electron capture are detectable by the current generation of X-ray spectrometers. While not as penetrating, these X-rays provides a useful complement to $\gamma$-rays, as in the detection of \isotope{44}{Ti} in G1.9+0.3 \citep{Borkowski2010} and the upper limit of \isotope{55}{Fe} in 1987A \citep{Leising2006}. }

In summary, understanding the production of radioactive \isotope{43}{K}, \isotope{47}{Ca}, \isotope{44,47}{Sc}, \isotope{44}{Ti}, \isotope{48,51}{Cr}, \isotope{48,49}{V}, \isotope{52,53}{Mn}, \isotope{55,59}{Fe}, \isotope{56,57}{Co}, and \isotope{56,57,59}{Ni} in CCSNe is important for the interpretation of past and future observations in terms of a broad range of CCSN physics, chemical evolution, and cosmic ray acceleration. The goal of the remainder of the paper is to identify the important nuclear reactions that need to be understood to make reliable predictions for the production of these isotopes during explosive silicon burning.

\section{Methods} \label{sec:methods}
\subsection{Supernova Model} \label{subsec:modeltraj}

We use thermodynamic trajectory data from a self-consistent \replaced{STIR}{Supernova Turbulence In Reduced-dimensionality (STIR)} explosion model of \citet{Couch2019} for a 12~$\Msun$ progenitor star from \citet{Sukhbold2016}.
The STIR model includes energy-dependent, two-moment neutrino transport in the ``M1'' approximation \citep{oconnor:2015}, a microphysical equation of state for dense matter \citep{steiner:2013}, and approximate general relativistic gravity \citep{marek:2006}. \added{Total energy is approximately conserved within the STIR model when accounting for diffusive mixing of energy and composition due to turbulent convection, as discussed in \citet{Couch2019, Warren2020}.}
The 1D explosion is achieved by a novel model for including the effects of convection and turbulence based on a Reynolds-decomposition of the hydrodynamic evolution equations which is then closed using the mixing length theory \citep{Couch2019}.
The explosion model we use here for a 12~$\Msun$ star results in a diagnostic explosion energy of $3.7\times10^{50}$~erg and a final baryonic proto-neutron star mass (i.e., mass cut) of 1.48 $\Msun$.
In STIR, both the explosion energy and proto-neutron star mass are predictions of the model, given a progenitor structure, and are not set by hand.

The nucleosynthesis is calculated using a post-processing approximation. Temperature ($T$), density ($\rho$), electron neutrino ($\nu_e$) flux, and electron anti-neutrino ($\bar{\nu}_e$) flux as functions of time (referred to as trajectories) are taken from 100 equal mass (\EE{4.537}{-3}~$M_\sun$ each) zones. These zones subdivide a range in stellar radius from 1.486~$M_\sun$ to 1.935~$M_\sun$ enclosed mass, comprising the silicon and oxygen shells of the progenitor as shown in Figure \ref{fig:initial_comp}. Figure \ref{fig:trajconditions} shows the thermodynamic characteristics of the trajectories using the following parameter definition for each trajectory: the peak temperature is the maximum temperature in GK, the total entropy is the entropy at peak temperature, which is nearly constant during the expansion, the peak radiation entropy is defined by \citet{Witti1994} as 
\begin{equation}
S_{\rm{rad}}(T_{9,\rm{peak}}) = 3.33\frac{T_{9,\rm{peak}}^3}{\rho_5(T_{9,\rm{peak}})}
\end{equation}
where $\rho_5 = 10^{-5}\times{\rho}$~g~cm$^{-3}$ with entropy in units of Boltzmann constant per baryon, and the peak density is the density at peak temperature in g~cm$^{-3}$. While the definition of peak radiation entropy assumes an ultra-relativistic electron-positron gas, the calculations in \skynet assume the gas is arbitrarily relativistic and degenerate. Additionally, the innermost trajectory has a neutron excess $\eta = 1-2Y_e = -0.015$ while all other trajectories are in the range $0.000 < \eta \leq 0.002$. Together these trajectories span an $\alpha$-rich freezeout parameter space similar to recent studies \citep{Magkotsios2010, Hoffman2010, Vance2020}. 

The region we consider is located just above the mass cut. Material closer to the star is assumed to fall back onto the compact remnant, while material at larger distances does not undergo sufficient shock heating to produce the medium mass $A=40$-60 nuclei of interest here. Since the supernova simulation ends at 1.93~s after bounce, the continuing evolution of the trajectories is modeled on an homologous expansion---the density scales as $\rho(t)=\rho_ft_f^3/t^3$, and the temperature uses the self-heating evolution described in \citet{Lippuner_2017}. We end the trajectories at 140~s after bounce, at which time all zones have cooled below 0.01 GK and no further nucleosynthesis occurs.

{
\begin{figure*}
\centering 
\plotone{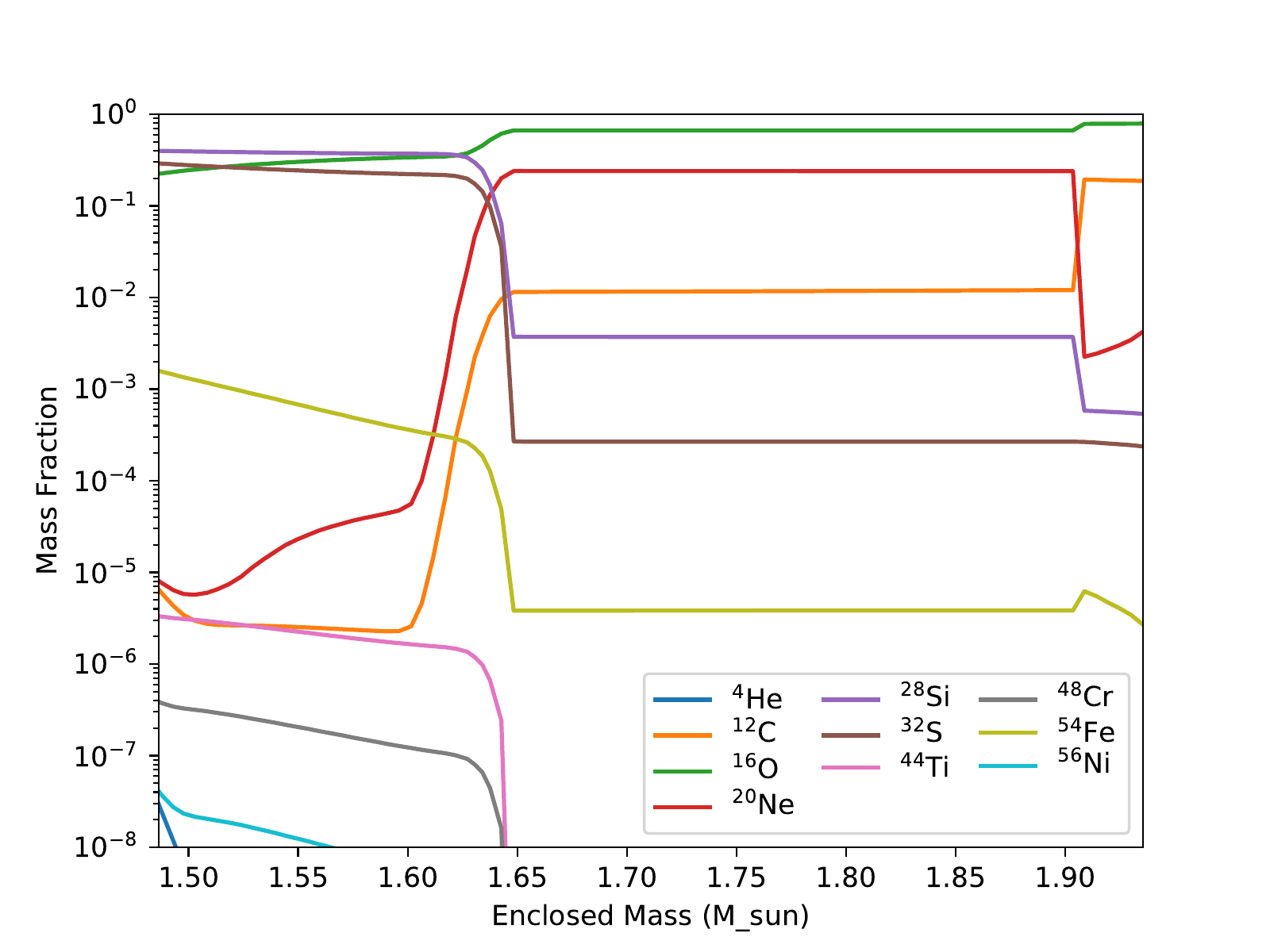}
\caption{Initial composition of notable isotopes of the 12~$M_\sun$ star before core-collapse. Only the simulated region of enclosed mass is shown.}
\label{fig:initial_comp}
\end{figure*}
}
{
\begin{figure}
\epsscale{1.20}
\plotone{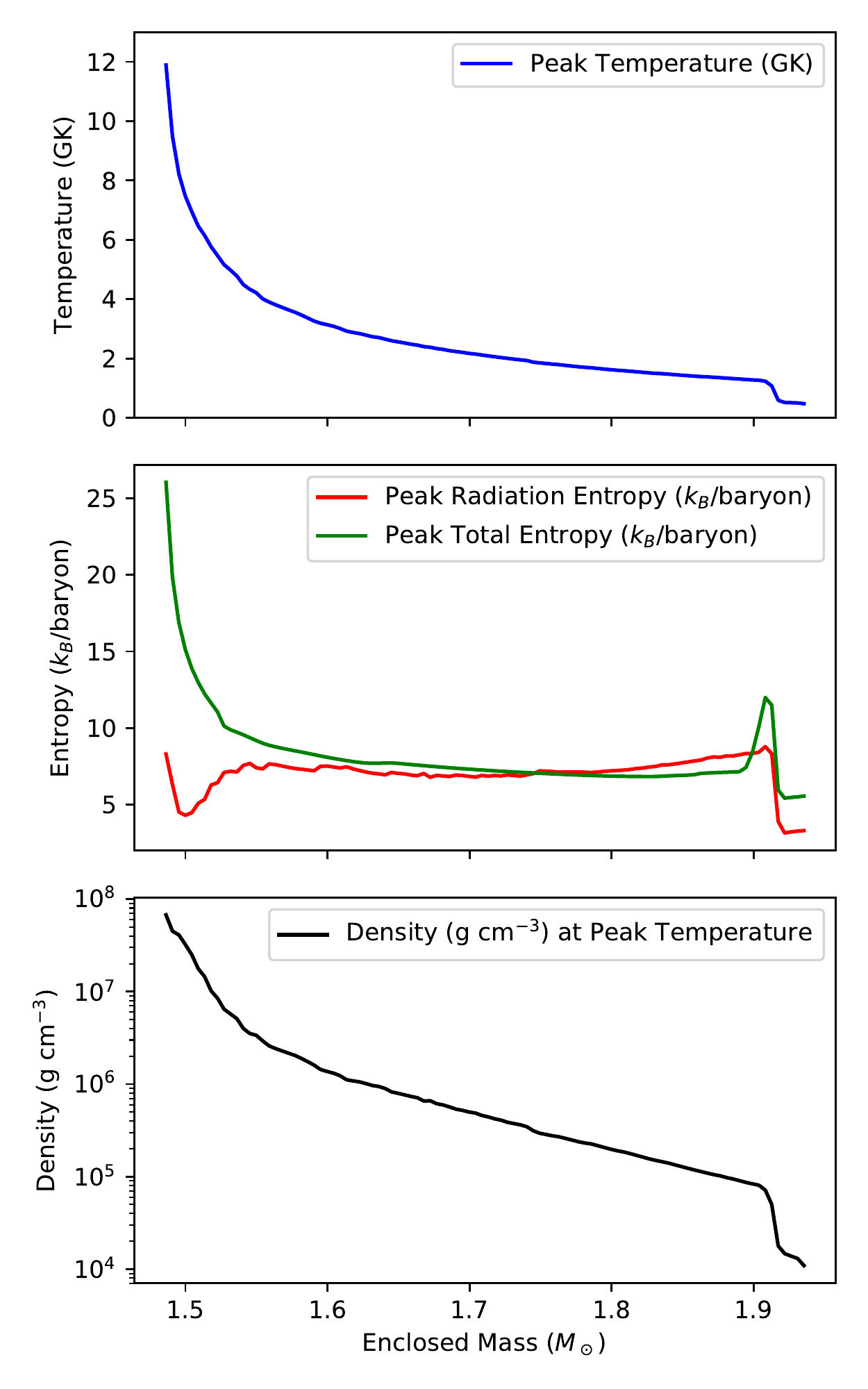}
\caption{Thermodynamic conditions for the simulated trajectories. Peak temperature (top), peak total and peak radiation entropy (middle), and peak density (bottom). For some trajectories, peak radiation entropy is larger than peak total entropy, this is due to approximations used in the definition of peak radiation entropy. }
\label{fig:trajconditions}
\end{figure}
}
{
\begin{figure*}
\centering 
\plotone{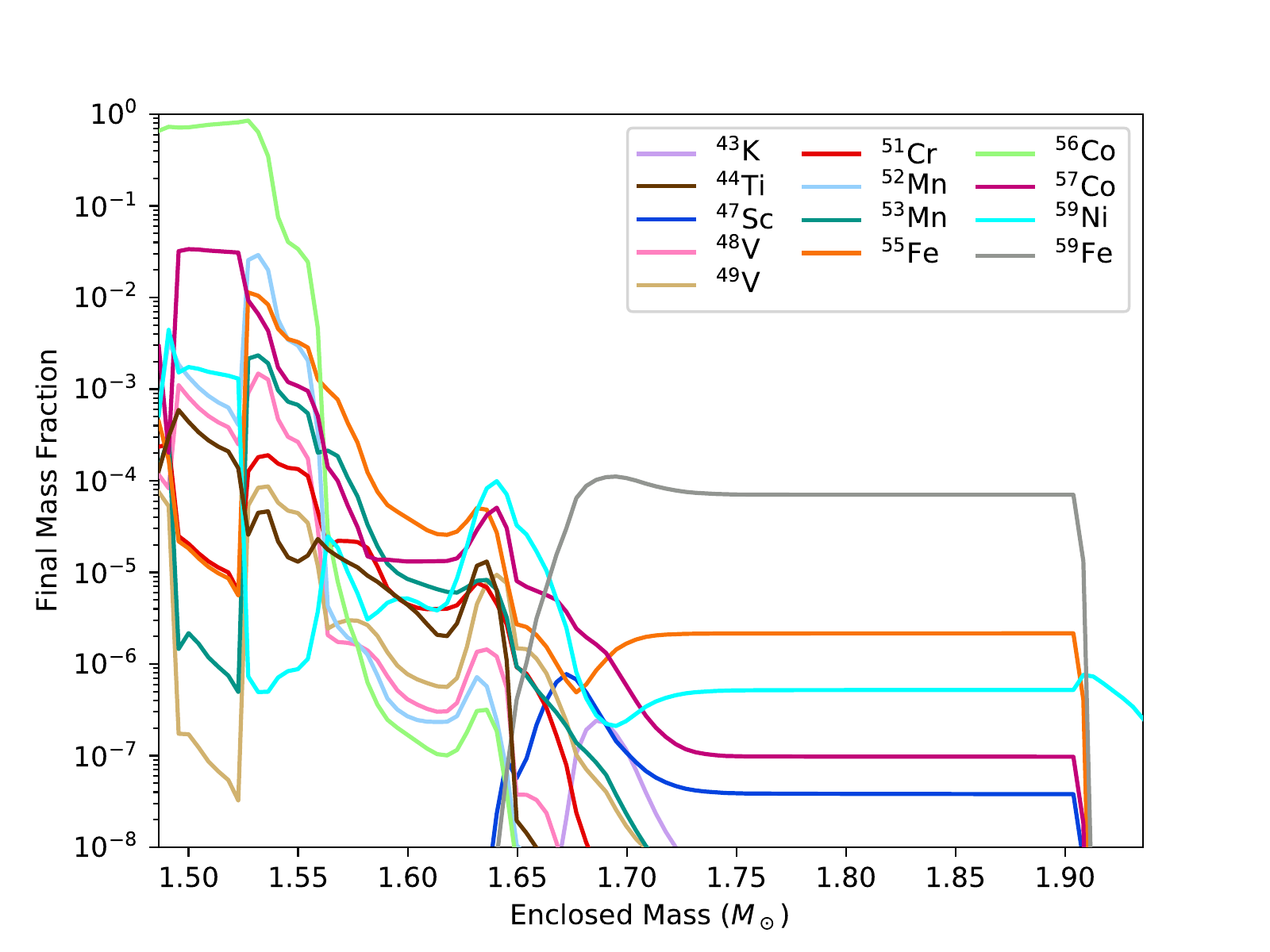}
\caption{Final mass fractions of isotopes of interest by enclosed mass shell for evolutions using standard REACLIB rates. Mass fractions include the short-lived parent isotopes in the mass chain (e.g. \isotope{56}{Co} includes both \isotope{56}{Co} and \isotope{56}{Ni}). Only the simulated region of enclosed mass is shown.}
\label{fig:production}
\end{figure*}
}

\subsection{Nuclear Reaction Network} \label{subsec:network}
For each of the 100 trajectories the nucleosynthesis and final composition are calculated using the nuclear reaction network library \skynet \citep{Lippuner_2017}. The network used here includes 1683 nuclides with element numbers $Z \leq 50$ and neutron numbers $N \leq 70$ connected by 22891 total reactions. Reactions considered include heavy ion fusion 
reactions; proton, neutron, and $\alpha$-induced reactions and their inverse; as well as $\beta$-decays, electron capture, and free nucleon-neutrino interactions. As a baseline set of nuclear reaction 
rates we use REACLIB v2.2 \citep{Cyburt_2010} and, where available, the weak reaction rates from \citet{Langanke2000, Oda1994}; and \citet{Fuller1985} as compiled by \citet{Paxton2015}. Free nucleon-neutrino interactions are included as described in \citet{Lippuner_2017} and include weak magnetism and recoil corrections from \citet{Burrows2006}, consistent with the rates used in the original CCSN simulation.

The initial composition of a trajectory is taken from the same 12~$\Msun$ progenitor model used for the supernova simulation \citep{Sukhbold2016} (see Figure \ref{fig:initial_comp}). 
The total ejected mass of an isotope of interest is determined by summing the contributions from all trajectories.  

The sensitivity of the produced abundance of an isotope of interest to nuclear reaction rates is determined by changing individual reaction rates one-by-one and re-calculating the nucleosynthesis for all trajectories
\citep{Iliadis2002}.  To reduce the number of reactions that need to be varied, we determined a subset of relevant reactions using the calculation with the baseline rates. Only reactions with a time integrated reaction flow above $10^{-10}$ in any trajectory were added to the list of varied reactions. Smaller reaction flows are negligible even when varying a reaction rate by a factor as large as 100 as the smallest isotopic abundances of interest are  $\approx10^{-8}$. This approach results in a subset of 3403 relevant reactions. Each of these reactions were individually varied up and down by factors of 100. To explore the linearity of the resulting abundance changes, we perform additional calculations for the 141 rates that significantly affect the synthesis of an isotope of interest, using smaller
variations of factors of 10 and 2. In total, 737,100 network evolutions were performed.

\begin{figure}
\epsscale{1.20}
\plotone{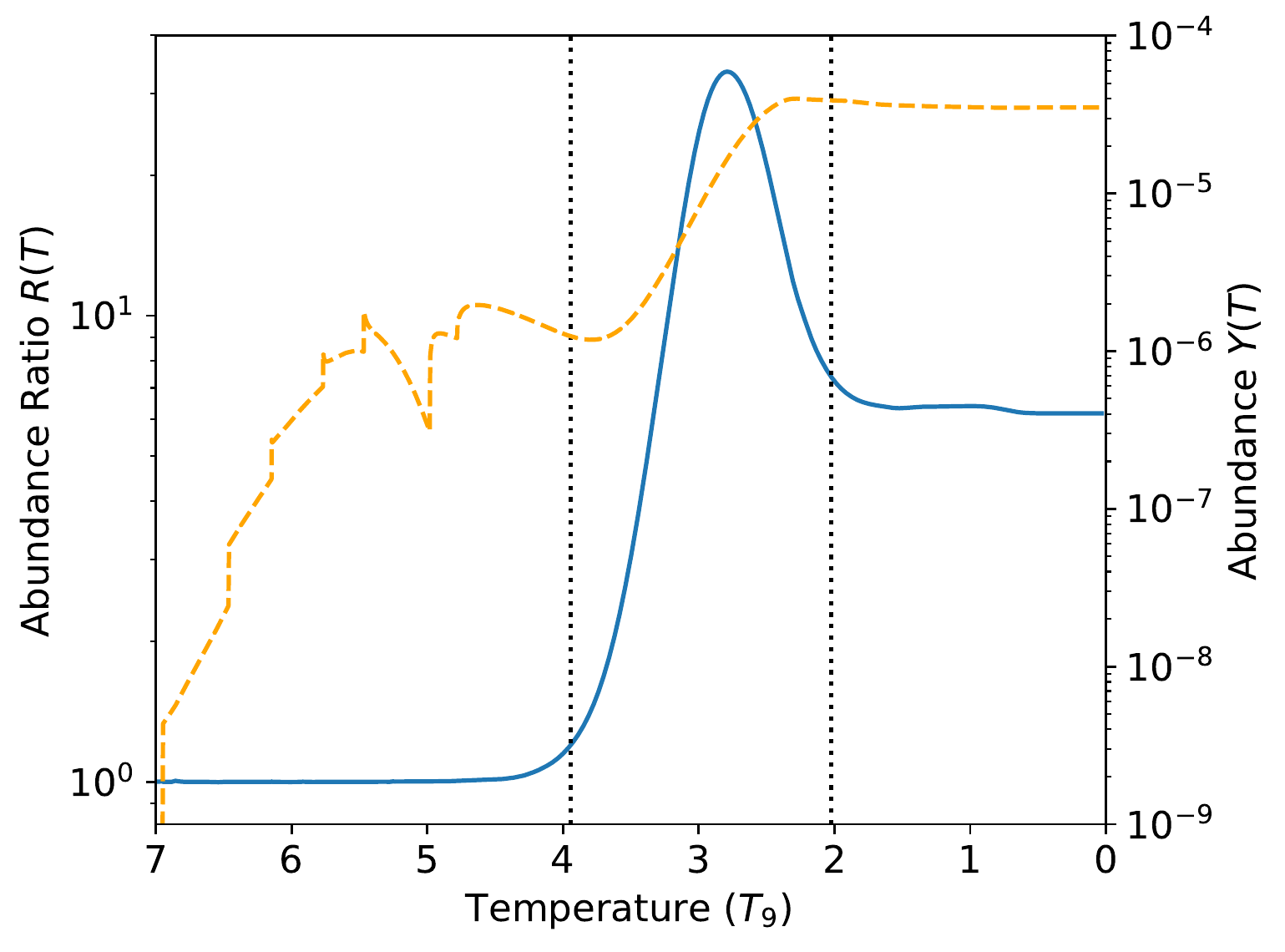}
\caption{$A=44$ abundance (orange dashed line) and ratio of $A=44$ abundance to the baseline abundance when varying \isotope{44}{Ti}\apreac \isotope{47}{V} by x0.01 as functions of temperature (blue line). Dotted black vertical lines indicate the determined temperature range where the reaction affects \isotope{44}{Ti} production.}
\label{fig:sample_ratio_plot}
\end{figure}

One goal of this work is to provide guidance for nuclear physics to improve the accuracy of nuclear reaction rates that affect the nucleosynthesis of isotopes of interest. For both experimental and theoretical nuclear physics work, it is critical to understand the temperature range over which a particular reaction rate needs to be determined. We developed an approach to obtain this information for each reaction rate that was identified to affect a specific isotope $i$. We take advantage of the fact that reaction rates only matter during the cooling of a trajectory once Nuclear Statistical Equilibrium (NSE) breaks down. At this late stage, in our model, the temperature is monotonically decreasing. Starting with the innermost and hottest trajectory, we sum together isotopic abundances from all trajectories with $T_{\rm{peak}}\geq T$ until we include all simulated trajectories. We can therefore uniquely determine at each temperature $T$ the sum of the abundances of the mass chain of isotope $i$ ($Y_i (T)$) over all zones contributing $\geq1\%$ to the total final abundance of that mass chain and calculate the ratio of the abundances obtained with the varied reaction rate ($Y'_i(T)$) to the baseline abundances,  
\begin{equation}
R_i(T)=Y'_i(T)/Y_i(T)
\end{equation}
Figure~\ref{fig:sample_ratio_plot} presents an example for the specific effect of varying \iso{Ti}{44}\apreac \iso{V}{47} by x0.01 on $A=44$ production. The sharp increases in mass chain abundance at specific, high temperatures are due to additional trajectories being included as the temperature drops. At high $T>6.5$~GK, $R_i(T)=1$ because abundances in NSE are insensitive to reaction rate changes. As the temperature drops, at some point $R_i$ starts deviating from 1 and transitions to the final value, $R_{f,i}$, at low temperatures when reactions freeze out completely. We determine the  temperature range $[T_{\rm min},T_{\rm max}]$ for the reaction rate sensitivity as 
the highest temperature where $\left|\log_{10} R_i(T_{\rm max})\right| \geq 10\% \left|\log_{10} R_{f,i} \right|$ and the lowest temperature where  $\left|\log_{10} R_i(T_{\rm min})\right| \leq 90\% \left|\log_{10} R_{f,i} \right|$. We ignore high temperature deviations of $R_i$ from 1 so long as $R_i$ returns to 1 for a sustained span of $0.05$~GK.

We emphasize that this is an approximate approach that makes a number of assumptions that are reasonable for this particular scenario. Most importantly we assume that temperature is dropping monotonically and that reaction rates continually become slower. The approach would not be appropriate for a scenario where some intermediate abundance is built up at one temperature, and is then processed further at another temperature (our method would only identify the second temperature range). The use of ratios instead of absolute abundances ensures that we are sensitive to reactions that change the abundance of the entire relevant equilibrium cluster even at times where the particular isotope of interest may not be fully populated within the cluster equilibrium. Also this method primarily narrows down the relevant temperature range, excluding sensitivity for higher and lower temperatures. Sensitivity within the given temperature range may not be evenly distributed. 

$[T_{\rm min},T_{\rm max}]$ then defines the temperature range over which the reaction rate has to be determined. This temperature range sensitively depends, even for the same reaction, on the isotope of interest. 

\section{Results} \label{sec:results}

Defining an impact factor,
\begin{equation}
F_i=10^{\left|\log_{10}R_{f,i} \right|}
\end{equation}
reaction rate variations that result in final abundance changes of $F_i\geq 1.1$ (effectively a 10\% difference) in an isotope of interest, $i$, are listed in Table~\ref{tab:effects} (moved to end for readability). Reactions are sorted by the mass number of the produced isotope of interest, and then in reverse order of impact defined as the maximum of $F_i$ for the x100 and x0.01 variations. Final abundances for the long-lived isotopes of interest, $Y_{f,{i}}$, were calculated by adding the abundances of parent isotopes that had not fully decayed yet at the end of the calculation. Decays of the isotopes of interest  are negligible within the 140~s of calculation time due to the much longer half-lives. For \iso{V}{48}, \iso{Co}{56}, and \iso{Co}{57} the parent isotopes are also long-lived isotopes of interest. In all three cases, these isotopes are predominantly produced by the decay of their parent isotope, with direct production being negligible at 0.1\%, $<0.01\%$, and 0.3\%, respectively. The reaction rate sensitivity of the parent abundance is therefore the same as the sensitivity for the daughter abundance. 

In total we find 141 reaction rates with $F_i\geq 1.1$ when varied by up to a factor of 100 for at least one isotope of interest $i$. The sensitivity of the final abundances to reaction rate variations varies widely. The lighter isotopes in the $A=43-49$ mass range are affected by a larger number of reaction rates. The most sensitive isotope is the lightest and most neutron-rich isotope studied here, $^{43}$K. When varied by a factor of a 100, 20 reaction rates have $F_i\geq2$, and 19 additional reaction rates have $F_i\geq 1.3$. For the other interesting isotopes in this mass range, 47 reaction rates have $F_i\geq1.3$ for at least one isotope $i$. On the other hand, for each of the isotopes in the $A=52-57$ mass range, there are no reaction rates with $F_i \geq 2$, and only 8 with $F_i\geq1.3$. The $A=52-57$ isotopes are closer to the peak of the NSE abundance distribution at $A=56$. Their synthesis is therefore more dominated by NSE, making them less susceptible to reaction rate variations. 

The most impactful reactions fall generally into two categories: reactions involving the isotope of interest or its parent nuclei, and rates which affect explosive nucleosynthesis more broadly. Reactions in the latter category are listed in Table~\ref{tab:broad_impact}. This group includes the  strongest sensitivity identified in this study, the impact of the \isotope{24}{Mg}\ngreac \isotope{25}{Mg} reaction on \isotope{43}{K} production.  This reaction affects  the production of all neutron-rich isotopes of interest. For the proton-rich isotopes, the reactions with the broadest impact are \triplealphanosp, \isotope{13}{N}\apreac \isotope{16}{O}, 
${}^{16}$O(${}^{12}$C,p)${}^{27}$Al, and \isotope{12}{C}\agreac \isotope{16}{O}, each of which significantly affects the production of 7 or more of the 13 isotopes of interest. The reactions are bottle-necks in the build-up of heavy elements during NSE freezeout. 

\begin{deluxetable}{cc}
\tablecaption{Reaction rate variations that affect three or more isotopes of interest \label{tab:broad_impact}}
\tablenum{2}

\tablehead{               Reaction& Number of Isotopes Affected}
\startdata
${}^{12}$C($\alpha$,$\gamma$)${}^{16}$O &    8 \\
$\alpha$($2\alpha$,$\gamma$)${}^{12}$C &    8 \\
${}^{16}$O(${}^{12}$C,p)${}^{27}$Al &    8 \\
${}^{13}$N($\alpha$,p)${}^{16}$O &    7 \\
${}^{27}$Al($\alpha$,n)${}^{30}$P &    6 \\
${}^{20}$Ne($\alpha$,$\gamma$)${}^{24}$Mg &    6 \\
${}^{44}$Ti($\alpha$,p)${}^{47}$V &    5 \\
${}^{42}$Ca($\alpha$,$\gamma$)${}^{46}$Ti &    5 \\
${}^{16}$O(${}^{12}$C,$\alpha$)${}^{24}$Mg &    5 \\
${}^{48}$Cr($\alpha$,p)${}^{51}$Mn &    4 \\
${}^{23}$Na($\alpha$,p)${}^{26}$Mg &    4 \\
${}^{53}$Fe(n,p)${}^{53}$Mn &    3 \\
${}^{52}$Fe($\alpha$,p)${}^{55}$Co &    3 \\
${}^{33}$S(n,$\alpha$)${}^{30}$Si &    3 \\
${}^{30}$Si(p,$\gamma$)${}^{31}$P &    3 \\
${}^{28}$Si(n,$\gamma$)${}^{29}$Si &    3 \\
${}^{28}$Al(p,$\alpha$)${}^{25}$Mg &    3 \\
${}^{27}$Si(n,${}^{12}$C)${}^{16}$O &    3 \\
${}^{27}$Al($\alpha$,p)${}^{30}$Si &    3 \\
${}^{26}$Mg($\alpha$,n)${}^{29}$Si &    3 \\
${}^{25}$Mg(p,$\gamma$)${}^{26}$Al &    3 \\
${}^{25}$Mg(n,$\gamma$)${}^{26}$Mg &    3 \\
${}^{25}$Mg($\alpha$,n)${}^{28}$Si &    3 \\
${}^{24}$Mg(n,$\gamma$)${}^{25}$Mg &    3 \\
${}^{20}$Ne(n,$\gamma$)${}^{21}$Ne &    3 \\
${}^{16}$O($\alpha$,$\gamma$)${}^{20}$Ne &    3 \\
\enddata
\end{deluxetable}

The other category of impactful reactions includes reactions which directly involve the isotope of interest or its parent nuclei. These include reactions like \isotope{42}{K}\ngreac \isotope{43}{K}, ${}^{47}$Sc(n,$\gamma$)${}^{48}$Sc, ${}^{49}$Mn(p,$\gamma$)${}^{50}$Fe, \isotope{51}{Mn}\pgreac \isotope{52}{Fe}, ${}^{52}$Fe($\alpha$,p)${}^{55}$Co, ${}^{53}$Fe(n,p)${}^{53}$Mn, ${}^{55}$Co(p,$\gamma$)${}^{56}$Ni, ${}^{57}$Ni(n,p)${}^{57}$Co, and \isotope{59}{Cu}\pgreac \isotope{60}{Zn}, which all are among the most impactful reactions for their respective isotopes, and for these isotopes only. By nature these reactions involve radioactive isotopes.  ${}^{44}$Ti($\alpha$,p)${}^{47}$V and ${}^{48}$Cr($\alpha$,p)${}^{51}$Mn are somewhat an exception falling into both categories. While they have 
the strongest impact on $^{44}$Ti and $^{51}$Cr production, respectively, they also impact a number of other isotopes of interest (see Table~\ref{tab:broad_impact}).


Table~\ref{tab:effects} also lists for each reaction rate variation the temperature range where the change in reaction rate produces the change in final abundance. This is the estimated temperature range over which the reaction rate needs to be known in order to predict nucleosynthesis reliably. These temperatures are mostly between 0.7~GK and 5.5~GK for all reactions. The \triplealpha reaction is an exception, often leading to changes at temperatures up to and above 6~GK.  The typical temperature ranges depend strongly on the isotope of interest. For $^{43}$K, $^{47}$Sc, $^{59}$Fe, and $^{59}$Ni the temperature range of interest is quite narrow (1.4 - 2.7~GK). The only outlier in that group is the $^{57}$Cu(p,$\gamma$)$^{58}$Zn reaction affecting $^{59}$Ni at relatively low temperatures of 0.8 - 1.1~GK.  For $^{44}$Ti, $^{48}$V, $^{49}$V, and $^{51}$Cr the temperature range of interest is much broader (mostly 0.7 - 5~GK). For $^{52}$Mn, $^{53}$Mn, $^{55}$Fe, and $^{57}$Co the temperature range is narrower but higher (2.0 - 5.5~GK) with two exceptions - the $^{52}$Fe(p,$\gamma$)$^{53}$Co reaction affects $^{53}$Mn at 0.73 - 1~GK, and the $^{56}$Ni(p,$\gamma$)$^{57}$Co reaction affects $^{57}$Co at very low temperatures of 0.47 - 0.65~GK. 



\section{Discussion} \label{sec:discussion}
\subsection{Nucleosynthesis}

Figure \ref{fig:nucleosynthesis} displays a comparison of the ejected masses of the mass chains of interest produced in the current model with two examples from previous work, \citet{Sukhbold2016} and \citet{Curtis2019}. Figure \ref{fig:ratio_to_solar} displays the ratios to solar abundances up to $A=80$. All three use a 12 $M_\sun$ progenitor, we take the progenitors evolved in \citet{Sukhbold2016}, which are based on \citet{Woosley2015}, while \citet{Curtis2019} uses \citet{Woosley2007} progenitors. In terms of nuclear physics, \citet{Curtis2019} also uses REACLIB reaction rates, while \citet{Sukhbold2016} uses older rates as in \citet{Woosley2007}.
The results from different models agree mostly within 40\%, except for $A=48$ where there is a 80\% difference between our result and the result of  \citet{Sukhbold2016}. While differences with respect to \citet{Sukhbold2016} could be in part due to use of different reaction rates, clearly there are also astrophysical uncertainties in the prediction of nucleosynthesis from explosive Si burning.

The $^{44}$Ti yield in this work of \EE{1.4}{-5}~$\Msun$ is about an order of magnitude lower than that inferred from observations of SN1987a or Cas A,  in line with previous 1D supernova model calculations (see Figure~\ref{fig:nucleosynthesis} and, for example, the summary in \citet{Chieffi2017}).

{
\begin{figure}
\epsscale{1.2}
\plotone{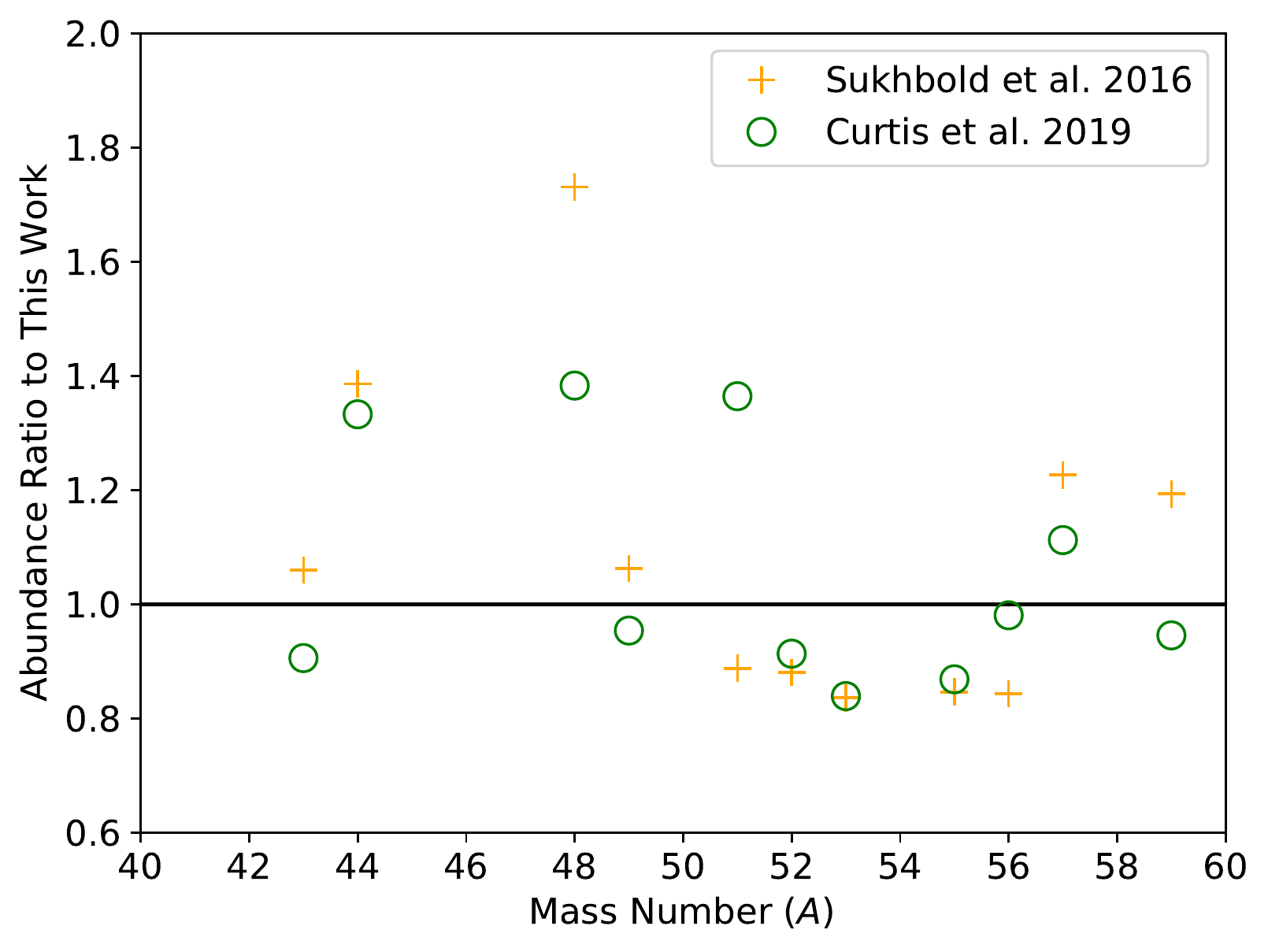}
\caption{Ratio of mass fraction of ejected isotopes in previous studies \citep{Sukhbold2016,Curtis2019} to this work, summed by mass number.}
\label{fig:nucleosynthesis}
\end{figure}
}
{
\begin{figure}
\epsscale{1.2}
\plotone{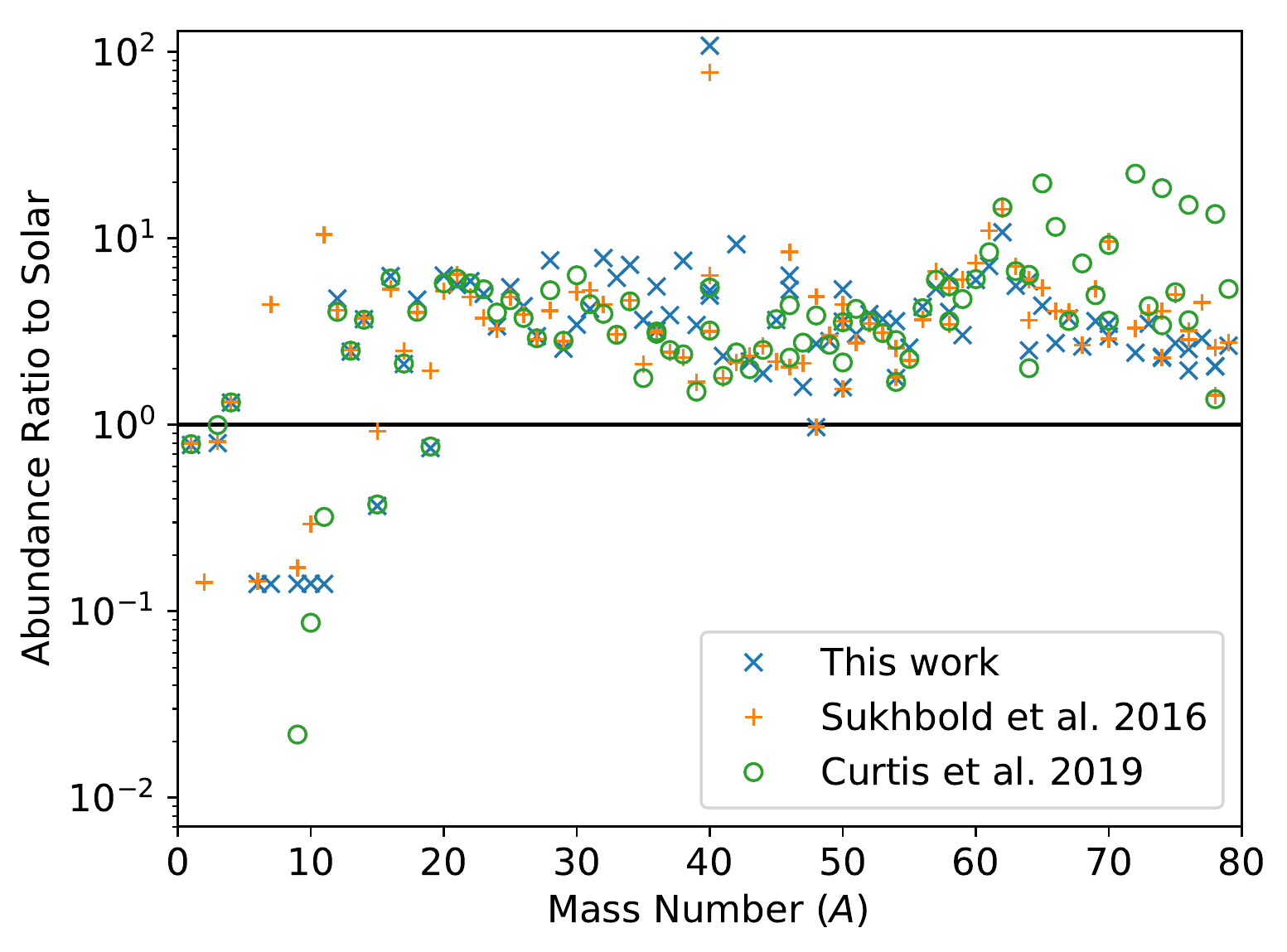}
\caption{Ratio of abundances of ejected isotopes to solar abundances summed by mass number.}
\label{fig:ratio_to_solar}
\end{figure}
}

\subsection{Previous Sensitivity Studies} \label{subsec:compare}
\subsubsection{\isotope{44}{Ti}}\label{subsubsec:44ti}
Given the importance of \isotope{44}{Ti} for $\gamma$-ray observations, a number of previous studies have identified critical nuclear reactions affecting its production, in particular the studies of \citet{The1998} and \citet{Magkotsios2010}. All studies, including this work, agree that $^{44}$Ti($\alpha$,p)$^{47}$V, $^{40}$Ca($\alpha$,$\gamma$)$^{44}$Ti and \triplealpha are the most important reactions governing the production of $^{44}$Ti. There is also agreement that the $^{45}$V(p,$\gamma$)$^{46}$Cr reactions plays a role, even though our model is much less sensitive to this reaction than both \citet{The1998} and \citet{Magkotsios2010}, where this reaction ranks near the top. On the other hand, we do find a significant sensitivity to $^{44}$Ti(p,$\gamma$)$^{45}$V, as does \citet{Magkotsios2010}, indicating that in our model (p,$\gamma$)-($\gamma$,p) equilibrium between $^{44}$Ti and $^{45}$V is not well established during $^{44}$Ti synthesis. 

For the remaining reactions identified in this work there are significant differences with  \citet{The1998}. Most of our sensitive reactions do not appear in their study, and we do not find the strong sensitivity to the \isotope{44}{Ti}($\alpha$,$\gamma$)\isotope{48}{Cr}, $^{57}$Ni(p,$\gamma$)$^{58}$Cu, and  $^{57}$Co(p,n)$^{57}$Ni reactions that they found. \citet{The1998} used a much narrower range of conditions and evolved a pure initial \isotope{28}{Si} composition from NSE conditions at $T=5.5$ GK and $\rho = 10^{7}$~g~cm${}^{-3}$ through adiabatic expansion. In contrast, our model utilizes several initial compositions and thermodynamic trajectories, only the first $\approx0.03$~$M_{\sun}$ of which achieve the NSE conditions of \citet{The1998}. Our model therefore includes both complete and incomplete explosive burning.
Even so, we do find a weak sensitivity on \isotope{44}{Ti}($\alpha$,$\gamma$)\isotope{48}{Cr} of $F_i\sim1.08$ (when varied by a factor of 100), just below our threshold.

Overall our results agree more closely with \citet{Magkotsios2010}, who scan a broad selection of peak temperatures and densities, indicating that more realistic models are needed to identify reaction rate sensitivities for $^{44}$Ti. There are only 4 out of 15 reactions in our list that do not appear in  \citet{Magkotsios2010}: $^{39}$K(p,$\alpha$)$^{36}$Ar, $^{27}$Al($\alpha$,n) $^{30}$P, $^{42}$Ca(p,$\alpha$)$^{39}$K, and $^{32}$S(n,$\alpha$)$^{30}$Si. However,  $^{39}$K(p,$\alpha$)$^{36}$Ar does appear prominently in \citet{The1998}. There are also three reactions that are ranked as ``Primary" in \citet{Magkotsios2010} that we do not find to have a strong sensitivity: \isotope{17}{F}($\alpha$,p)\isotope{20}{Ne}, \isotope{21}{Na}($\alpha$,p)\isotope{24}{Mg}, and again  \isotope{57}{Ni}(p,$\gamma$)\isotope{58}{Cu}.  \isotope{57}{Ni}(p,$\gamma$)\isotope{58}{Cu} has $F_i \sim 1.03$, just below our threshold. They find the other two reactions to be significant only in regions of higher peak densities ($\rho\sim 10^{9}$~g~cm$^{-3}$) than our conditions.  

\subsubsection{\isotope{55}{\textrm{Fe}}, \isotope{57}{Co}, \isotope{59}{Ni}}\label{subsubsec:55fe57co59ni}
Reactions relevant for \isotope{55}{Fe}, \isotope{57}{Co}, and \isotope{59}{Ni} production were identified in \citet{Jordan2003}, who use a similar exponential expansion model as \citet{The1998}. For \isotope{55}{Fe}, we find only one of the four rates identified in \citet{Jordan2003}, \isotope{55}{Co}\pgreac \isotope{56}{Ni}, to be significant. Of the three others, \triplealpha is found to cause a $F_i\sim1.09$ effect at a factor of x100 change, and both \isotope{59}{Cu}\pareac \isotope{56}{Ni}, and \isotope{59}{Cu}\pgreac \isotope{60}{Zn} have only a minimal effect. As before, the simplified model and the more limited range of conditions explored in \citet{Jordan2003} are likely the reason for the discrepancy. Much of the \isotope{55}{Fe} production in our model occurs in the trajectories which do not undergo complete burning and do not reach the temperatures or densities specified in \citet{Jordan2003}. Because of this, we generate a higher mass fraction ($\sim$~x10) of \isotope{55}{Fe} than \citet{Jordan2003}, and obtain a more complete picture of \isotope{55}{Fe} production. For \isotope{57}{Co} and \isotope{59}{Ni} our results agree with \citet{Jordan2003} on the most important reactions, \isotope{57}{Ni}(n,p)\isotope{57}{Co} in the case of  \isotope{57}{Co} and \isotope{59}{Cu} \pgreac \isotope{60}{Zn}, \isotope{59}{Cu}\pareac \isotope{56}{Ni} and \triplealpha in the case of  \isotope{59}{Ni}. For all three isotopes we explore not only x10 variations as in \citet{Jordan2003}, but also x100 variations. Consequently we identify a number of additional relevant  reactions. 

\subsection{Rate Uncertainties} \label{subsec:rate_uncertainty}
The main goal of this paper is to identify the nuclear reaction rates that determine long-lived radioisotope production in CCSNe. However, to provide guidance for nuclear physics on which of these reaction rates need improved accuracy, their current uncertainties must be considered. Such uncertainties are often difficult to estimate reliably, especially in the case of theoretical predictions. Nevertheless, to provide some approximate guidance, we used the uncertainty factors listed in the current STARLIB database \citep{Sallaska_2013} for the relevant temperature ranges that we identified in this work. While past studies \citep{Fields2018} used Monte Carlo methods to sample STARLIB rates within the prescribed uncertainties, here we take a more simplified approach. We interpolated our sensitivities in Table~\ref{tab:effects} and determined the expected final abundance changes for rate variations within the STARLIB 1$\sigma$ uncertainty (the impact factor). We considered both rate increases and rate decreases and used the larger sensitivity. Table~\ref{tab:10x} provides a list of reactions ordered by their impact factor, $F_i$, for $F_i \geq 1.1$. 16 reactions have $F_i\geq 1.5$. Reactions with large impact factors indicate the most important reaction rate uncertainties to be addressed for improved nucleosynthesis predictions.  

We emphasize that the impact factors provide only some approximate guidance. In addition to astrophysical uncertainties and possible correlations between rate uncertainties, there are large uncertainties in the estimates of the reaction rate uncertainties. For example, the \triplealpha reaction does not appear in this list despite its importance, as its uncertainty in STARLIB is less than 35\%. However, this rate needs to be known up to a temperature of around 5.8~GK. At such high temperatures, the role of resonances above the Hoyle state in $^{12}$C remains unclear and uncertainties are likely underestimated considerably \citep{Zimmerman2013}. In addition, it has been pointed out, that proton induced scattering at sufficiently high densities and temperatures can lead to orders of magnitude increases in the \triplealpha reaction rate \citep{Beard:2017}. The impact of these effects, and their uncertainties, needs to be investigated. The sensitivities provided in Table \ref{tab:effects} can then be used to determine whether an improved reaction rate is needed for explosive Si burning nucleosynthesis. 

\startlongtable
\begin{deluxetable}{lcc}
\tabletypesize{\footnotesize}
\tablecaption{Reaction rates with the largest impact based on their current estimated uncertainties.}
\tablenum{3}

\tablehead{Reaction & Impact & Isotope Affected }
\startdata
${}^{42}$K(n,$\gamma$)${}^{43}$K &  4.18 & \iso{K }{ 43}\\
${}^{44}$Ti($\alpha$,p)${}^{47}$V &  2.61, 1.31, 1.12\tablenotemark{a} & \iso{Ti}{ 44}, \iso{V}{48}, \iso{V}{49}\\
${}^{43}$K(p,n)${}^{43}$Ca &  2.51 & \iso{K }{ 43}\\
${}^{59}$Cu(p,$\gamma$)${}^{60}$Zn &  2.16 & \iso{Ni}{ 59}\\
${}^{42}$K(p,n)${}^{42}$Ca &  2.13 & \iso{K }{ 43}\\
${}^{23}$Na($\alpha$,p)${}^{26}$Mg &  2.12, 1.14, 1.13, 1.12\tablenotemark{a} & \iso{K }{ 43}, \iso{Sc}{47} \iso{V }{  49}, \iso{Fe}{55}\\
${}^{27}$Al($\alpha$,p)${}^{30}$Si &  1.91, 1.58\tablenotemark{a} & \iso{K }{ 43}, \iso{Sc}{47}\\
${}^{28}$Al(p,$\alpha$)${}^{25}$Mg &  1.89, 1.37\tablenotemark{a} & \iso{K }{ 43}, \iso{Sc}{47}\\
${}^{47}$Sc(n,$\gamma$)${}^{48}$Sc &  1.88 & \iso{Sc}{ 47}\\
${}^{47}$Ti(n,p)${}^{47}$Sc &  1.85 & \iso{Sc}{ 47}\\
${}^{48}$Cr($\alpha$,p)${}^{51}$Mn &  1.84, 1.16\tablenotemark{a} & \iso{V }{ 48}, \iso{Cr}{51}\\
${}^{51}$Mn(p,$\gamma$)${}^{52}$Fe &  1.76 & \iso{Cr}{ 51}\\
${}^{41}$K(p,$\alpha$)${}^{38}$Ar &  1.72 & \iso{K }{ 43}\\
${}^{43}$K(n,$\gamma$)${}^{44}$K &  1.65 & \iso{K }{ 43}\\
${}^{46}$Sc(n,$\gamma$)${}^{47}$Sc &  1.55 & \iso{Sc}{ 47}\\
${}^{46}$Sc(p,n)${}^{46}$Ti &  1.45 & \iso{Sc}{ 47}\\
${}^{53}$Fe(n,p)${}^{53}$Mn &  1.41 & \iso{Mn}{ 53}\\
${}^{49}$Mn(p,$\gamma$)${}^{50}$Fe &  1.34 & \iso{V }{ 49}\\
${}^{55}$Co(p,$\gamma$)${}^{56}$Ni &  1.32 & \iso{Fe}{ 55}\\
${}^{45}$Ca(n,$\gamma$)${}^{46}$Ca &  1.31 & \iso{Sc}{ 47}\\
${}^{32}$S(n,$\alpha$)${}^{29}$Si &  1.31, 1.29\tablenotemark{a} & \iso{K }{ 43}, \iso{Sc}{47}\\
${}^{40}$Ar(p,$\gamma$)${}^{41}$K &  1.30 & \iso{K }{ 43}\\
${}^{44}$Ca(p,$\gamma$)${}^{45}$Sc &  1.29 & \iso{Sc}{ 47}\\
${}^{40}$K(n,$\gamma$)${}^{41}$K &  1.29 & \iso{K }{ 43}\\
${}^{45}$Sc(p,$\gamma$)${}^{46}$Ti &  1.27 & \iso{Sc}{ 47}\\
${}^{59}$Cu(p,$\alpha$)${}^{56}$Ni &  1.25 & \iso{Ni}{ 59}\\
${}^{49}$Cr(n,p)${}^{49}$V &  1.25 & \iso{V }{ 49}\\
${}^{57}$Ni(n,p)${}^{57}$Co &  1.24, 1.21\tablenotemark{a} & \iso{Co}{ 57}, \iso{Ni}{59}\\
${}^{41}$Ca(n,$\alpha$)${}^{38}$Ar &  1.23 & \iso{K }{ 43}\\
${}^{41}$K(p,n)${}^{41}$Ca &  1.21 & \iso{K }{ 43}\\
${}^{59}$Fe(n,$\gamma$)${}^{60}$Fe &  1.19 & \iso{Fe}{ 59}\\
${}^{49}$V(p,$\gamma$)${}^{50}$Cr &  1.19 & \iso{V }{ 49}\\
${}^{25}$Mg($\alpha$,n)${}^{28}$Si &  1.19, 1.11\tablenotemark{a} & \iso{Sc}{ 47}, \iso{K}{43}\\
${}^{43}$Sc(p,$\gamma$)${}^{44}$Ti &  1.18 & \iso{Ti}{ 44}\\
${}^{57}$Cu(p,$\gamma$)${}^{58}$Zn &  1.17 & \iso{Ni}{ 59}\\
${}^{46}$Ca(p,$\gamma$)${}^{47}$Sc &  1.16 & \iso{Sc}{ 47}\\
${}^{52}$Fe($\alpha$,p)${}^{55}$Co &  1.15 & \iso{Mn}{ 52}\\
${}^{48}$Cr(p,$\gamma$)${}^{49}$Mn &  1.15 & \iso{V }{ 49}\\
${}^{43}$Sc(p,$\alpha$)${}^{40}$Ca &  1.15 & \iso{Ti}{ 44}\\
${}^{41}$Ca(n,$\gamma$)${}^{42}$Ca &  1.15 & \iso{Sc}{ 47}\\
${}^{39}$Ar(n,$\gamma$)${}^{40}$Ar &  1.15 & \iso{K }{ 43}\\
${}^{28}$Al(p,n)${}^{28}$Si &  1.15 & \iso{K }{ 43}\\
${}^{13}$N($\alpha$,p)${}^{16}$O &  1.15, 1.15, 1.14, 1.14, 1.11\tablenotemark{a} & \iso{Mn}{ 52}, \iso{Fe}{55}, \iso{V}{49}, \iso{Mn}{53}, \iso{V}{48}\\
${}^{49}$Cr(p,$\gamma$)${}^{50}$Mn &  1.14 & \iso{V }{ 49}\\
${}^{40}$Ca($\alpha$,$\gamma$)${}^{44}$Ti &  1.14 & \iso{Ti}{ 44}\\
${}^{41}$K(p,$\gamma$)${}^{42}$Ca &  1.13 & \iso{K }{ 43}\\
${}^{47}$Ca(p,n)${}^{47}$Sc &  1.11 & \iso{Sc}{ 47}\\
${}^{45}$V(p,$\gamma$)${}^{46}$Cr &  1.10 & \iso{Ti}{ 44}\\
\enddata
		\label{tab:10x}
\tablenotetext{a}{For reactions that impact multiple isotopes of interest, the impact numbers correspond to the isotopes listed, respectively.}
\end{deluxetable}




\section{Conclusion} \label{sec:conclusion}
We determined the sensitivity of the production of long-lived radioactive isotopes in a model for explosive Si burning in a CCSN. We vary individual reaction rates to determine the local derivative of the final abundance of a given isotope with respect to a given reaction rate. This approach has been used in previous work (e.g., \citet{Iliadis2002}) and enables the identification of critical reaction rates without making assumptions about their uncertainties. Compared to previous studies, we investigate the synthesis of a much broader range of 13 radioisotopes from \isotope{43}{K} to \isotope{59}{Ni} using a new model for 1D CCSN explosions that incorporates the crucial effects of convection and turbulence and does not resort to altering the microphysics to achieve explosions \citep{Couch2019}. We also developed a new method to identify the relevant temperature range for each reaction, and find that these temperature ranges depend sensitively on the reaction and final isotope of interest.
Our results can be used to determine the importance of a given reaction rate uncertainty (in the relevant temperature range provided by our work) for CCSN explosive nucleosynthesis predictions, and provide guidance on which reactions require further study, both experimental and theoretical. This is of particular importance in light of new emerging capabilities in nuclear experiment and nuclear theory. To provide some initial guidance in this direction, we use the reaction rate uncertainties provided by the STARLIB database to produce a ranked list of the most important reaction rate uncertainties. We emphasize however that for a final decision on the importance of a reaction rate a thorough analysis of the possible uncertainties is required, as well as considerations of the individual accuracy needs for the various long-lived radioactive isotopes.

The future reduction of nuclear physics uncertainties in explosive Si burning nucleosynthesis models enabled by our work will be important to prepare the field for advances in X- and $\gamma$-ray observations \citep{Timmes2019}, in stardust analysis \citep{Stephan2016}, and for the potential observation of a galactic supernova. Reduced and well characterized nuclear physics uncertainties will enable quantitative comparisons of astrophysical models with observations. While our results for $^{44}$Ti are in overall reasonable agreement  with the one previous study \citep{Magkotsios2010} that explored a similar realistic range of conditions, there are some differences, especially for the weaker sensitivities. This may indicate that a broader range of sensitivity studies for different astrophysical models may be needed to identify the individual nuclear physics needs of each model. 




\acknowledgments
We would like to thank Tuguldur Sukhbold for providing the full initial isotopic compositions used as the presupernova stellar model and Evan O'Connor for helpful discussions.
This work was supported by the National Science Foundation under award numbers PHY-1102511, PHY-1913554, PHY-1430152 (JINA Center for the Evolution of the Elements), and in part through computational resources and services provided by the Institute for Cyber-Enabled Research at Michigan State University.
SMC is supported by the U.S. Department of Energy, Office of Science, Office of Nuclear Physics, under Award Numbers DE-SC0015904 and DE-SC0017955. MLW is supported by an NSF Astronomy and Astrophysics Postdoctoral Fellowship under award AST-1801844.

\added{\software{matplotlib \citep{matplotlib}, NumPy \citep{numpy}, pandas \citep{Mckinney2010}, \skynet \citep{Lippuner_2017}}}

\startlongtable
\begin{deluxetable*}{lll|cccccc|cc}
\tablecaption{Final Abundance Changes Resulting from Reaction-Rate Variations \label{tab:effects}}
\tablenum{1}

\tablehead{&&&         \multicolumn6c{$R_f$ For Reaction Rate Multiplied By:}         &         \multicolumn{2}{|c}{$T_9$ Range} \\
            \cline{4-9} \cline{10-11}
              Isotope & $X_{i,\rm rec}$ & Reaction & 100 & 10 & 2 & 0.5 & 0.1 & 0.01 & $T_{9,\rm low}$& $T_{9,\rm high}$
}
\startdata
\hline${}^{43}$K&1.4e-08&${}^{24}$Mg(n,$\gamma$)${}^{25}$Mg&0.139&0.196&0.571&1.61&3.12&4.01&1.39&2.38\\
&&${}^{42}$K(n,$\gamma$)${}^{43}$K&3.97&2.93&1.54&0.622&0.239&0.139&1.39&2.38\\
&&${}^{25}$Mg($\alpha$,n)${}^{28}$Si&6.94&4.47&1.62&0.697&0.484&0.443&2.15&2.4\\
&&${}^{25}$Mg(n,$\gamma$)${}^{26}$Mg&0.166&0.333&0.764&1.2&1.47&1.55&1.41&2.38\\
&&${}^{43}$K(p,n)${}^{43}$Ca&0.172&0.398&0.771&1.26&1.95&2.78&1.42&2.6\\
&&${}^{20}$Ne(n,$\gamma$)${}^{21}$Ne&0.2&0.51&0.892&1.07&1.12&1.14&1.44&2.45\\
&&${}^{42}$K(p,n)${}^{42}$Ca&0.203&0.47&0.828&1.16&1.39&1.49&1.44&2.45\\
&&${}^{25}$Mg(p,$\gamma$)${}^{26}$Al&0.222&0.452&0.872&1.08&1.17&1.19&1.46&2.4\\
&&${}^{41}$K(n,$\gamma$)${}^{42}$K&3.62&2.79&1.49&0.666&0.333&0.247&1.47&2.38\\
&&${}^{28}$Al(p,$\alpha$)${}^{25}$Mg&4.04&1.89&1.12&0.941&0.894&0.884&2.14&2.4\\
&&${}^{23}$Na($\alpha$,p)${}^{26}$Mg&0.259&0.471&0.821&1.16&1.4&1.48&1.47&2.38\\
&&${}^{27}$Al($\alpha$,p)${}^{30}$Si&0.294&0.523&0.857&1.12&1.27&1.32&2.06&2.4\\
&&${}^{20}$Ne($\alpha$,$\gamma$)${}^{24}$Mg&0.3&0.525&0.855&1.12&1.26&1.31&2.1&2.6\\
&&${}^{41}$K(p,$\alpha$)${}^{38}$Ar&0.313&0.581&0.884&1.09&1.19&1.22&2.09&2.38\\
&&${}^{43}$K(n,$\gamma$)${}^{44}$K&0.315&0.605&0.911&1.06&1.11&1.12&2.11&2.38\\
&&${}^{28}$Si(n,$\gamma$)${}^{29}$Si&0.322&0.742&0.954&1.03&1.05&1.06&2.17&2.38\\
&&${}^{26}$Mg($\alpha$,n)${}^{29}$Si&2.94&2.29&1.35&0.76&0.539&0.489&2.15&2.38\\
&&${}^{27}$Al(p,$\gamma$)${}^{28}$Si&2.53&1.41&1.05&0.972&0.949&0.943&2.14&2.48\\
&&${}^{27}$Al(n,$\gamma$)${}^{28}$Al&2.02&1.3&1.04&0.981&0.967&0.964&2.1&2.48\\
&&${}^{29}$Si(n,$\gamma$)${}^{30}$Si&0.499&0.797&0.962&1.02&1.04&1.05&2.17&2.38\\
&&${}^{24}$Mg($\alpha$,$\gamma$)${}^{28}$Si&1.93&1.14&1.02&0.991&0.985&0.984&2.09&2.48\\
&&${}^{23}$Na(p,$\alpha$)${}^{20}$Ne&1.09&1.08&1.04&0.935&0.721&0.53&2.15&2.38\\
&&${}^{30}$Si(p,$\gamma$)${}^{31}$P&1.82&1.42&1.07&0.962&0.93&0.923&2.11&2.56\\
&&${}^{41}$K(p,$\gamma$)${}^{42}$Ca&0.55&0.886&0.984&1.01&1.01&1.02&2.12&2.38\\
&&${}^{21}$Ne($\alpha$,n)${}^{24}$Mg&1.07&1.06&1.03&0.95&0.764&0.582&2.24&2.38\\
&&${}^{27}$Al(p,$\alpha$)${}^{24}$Mg&0.762&0.796&0.904&1.13&1.44&1.67&2.14&2.48\\
&&${}^{29}$Si(p,$\gamma$)${}^{30}$P&0.621&0.832&0.968&1.02&1.04&1.04&2.14&2.38\\
&&${}^{40}$Ar(p,$\gamma$)${}^{41}$K&1.6&1.3&1.07&0.951&0.899&0.885&2.13&2.38\\
&&${}^{32}$S(n,$\alpha$)${}^{29}$Si&1.59&1.31&1.06&0.969&0.944&0.938&2.17&2.4\\
&&${}^{40}$K(n,$\gamma$)${}^{41}$K&1.57&1.29&1.06&0.968&0.94&0.933&2.15&2.38\\
&&${}^{40}$Ca(n,$\gamma$)${}^{41}$Ca&1.54&1.31&1.07&0.957&0.918&0.909&2.15&2.38\\
&&${}^{21}$Ne(p,$\gamma$)${}^{22}$Na&0.71&0.925&0.992&1.0&1.01&1.01&2.22&2.38\\
&&${}^{41}$K(p,n)${}^{41}$Ca&1.37&1.21&1.04&0.974&0.95&0.945&2.27&2.38\\
&&${}^{28}$Al(p,n)${}^{28}$Si&1.04&1.03&1.02&0.973&0.869&0.747&2.14&2.38\\
&&${}^{41}$Ca(n,$\alpha$)${}^{38}$Ar&0.893&0.905&0.953&1.07&1.23&1.33&2.15&2.38\\
&&${}^{39}$Ar(n,$\gamma$)${}^{40}$Ar&1.32&1.15&1.03&0.986&0.973&0.97&2.15&2.38\\
&&${}^{39}$K(n,$\gamma$)${}^{40}$K&1.31&1.19&1.05&0.971&0.944&0.937&2.17&2.38\\
&&${}^{26}$Mg(p,$\gamma$)${}^{27}$Al&1.14&1.29&1.11&0.909&0.797&0.763&2.06&2.38\\
&&${}^{37}$Cl($\alpha$,$\gamma$)${}^{41}$K&1.3&1.03&1.0&0.998&0.997&0.997&2.17&2.38\\
&&${}^{16}$O($\alpha$,$\gamma$)${}^{20}$Ne&1.07&1.05&1.02&0.96&0.905&0.803&2.02&2.6\\
&&${}^{27}$Al($\alpha$,n)${}^{30}$P&0.803&0.957&0.994&1.0&1.01&1.01&2.11&2.38\\
&&${}^{16}$O(${}^{12}$C,p)${}^{27}$Al&0.806&0.975&0.997&1.0&1.0&1.0&2.14&2.38\\
&&${}^{32}$S(n,$\gamma$)${}^{33}$S&0.819&0.899&0.977&1.02&1.03&1.04&2.21&2.38\\
&&${}^{12}$C(${}^{12}$C,$\alpha$)${}^{20}$Ne&1.21&1.04&1.0&0.997&0.995&0.995&1.98&2.38\\
&&${}^{29}$Si($\alpha$,$\gamma$)${}^{33}$S&0.841&0.976&0.997&1.0&1.0&1.0&2.15&2.38\\
&&${}^{38}$Ar(p,$\gamma$)${}^{39}$K&1.18&1.07&1.01&0.992&0.985&0.983&2.15&2.38\\
&&${}^{42}$K(p,$\alpha$)${}^{39}$Ar&0.848&0.982&0.998&1.0&1.0&1.0&2.12&2.38\\
&&${}^{30}$Si(n,$\gamma$)${}^{31}$Si&1.15&1.03&1.0&0.998&0.997&0.997&2.1&2.4\\
&&${}^{39}$Ar(p,$\gamma$)${}^{40}$K&1.15&1.04&1.01&0.997&0.995&0.994&2.15&2.38\\
&&${}^{22}$Ne($\alpha$,n)${}^{25}$Mg&1.02&1.02&1.01&0.981&0.916&0.874&2.17&2.38\\
&&${}^{28}$Al(n,$\gamma$)${}^{29}$Al&0.878&0.978&0.997&1.0&1.0&1.0&2.21&2.38\\
&&${}^{23}$Na($\alpha$,$\gamma$)${}^{27}$Al&0.885&0.986&0.998&1.0&1.0&1.0&2.17&2.38\\
&&${}^{21}$Ne(n,$\gamma$)${}^{22}$Ne&0.893&0.987&0.998&1.0&1.0&1.0&2.24&2.38\\
&&${}^{26}$Al(n,p)${}^{26}$Mg&0.922&0.933&0.97&1.03&1.1&1.12&2.15&2.38\\
&&${}^{43}$K(p,$\alpha$)${}^{40}$Ar&0.901&0.989&0.999&1.0&1.0&1.0&2.1&2.38\\
&&${}^{41}$Ca(n,$\gamma$)${}^{42}$Ca&0.904&0.962&0.994&1.0&1.01&1.01&2.15&2.38\\
&&${}^{40}$K(n,$\alpha$)${}^{37}$Cl&0.907&0.924&0.971&1.03&1.06&1.07&2.15&2.38\\
&&${}^{23}$Na(n,$\gamma$)${}^{24}$Na&0.908&0.97&0.996&1.0&1.0&1.0&2.05&2.38\\
&&${}^{38}$Ar(n,$\gamma$)${}^{39}$Ar&1.1&1.06&1.01&0.991&0.983&0.981&2.12&2.38\\
\hline${}^{44}$Ti&3.01e-05&${}^{44}$Ti($\alpha$,p)${}^{47}$V&0.209&0.385&0.74&1.35&2.61&5.74&0.88&3.94\\
&&$\alpha$($2\alpha$,$\gamma$)${}^{12}$C&1.3&1.0&1.01&0.929&0.528&0.175&1.83&6.15\\
&&${}^{40}$Ca($\alpha$,$\gamma$)${}^{44}$Ti&1.96&1.62&1.19&0.83&0.543&0.374&0.97&3.21\\
&&${}^{12}$C($\alpha$,$\gamma$)${}^{16}$O&2.15&1.2&1.02&0.988&0.977&0.975&2.1&4.98\\
&&${}^{43}$Sc(p,$\gamma$)${}^{44}$Ti&1.35&1.18&1.04&0.977&0.958&0.951&1.66&3.09\\
&&${}^{13}$N($\alpha$,p)${}^{16}$O&1.29&1.05&1.01&0.994&0.992&1.01&2.08&4.98\\
&&${}^{43}$Sc(p,$\alpha$)${}^{40}$Ca&0.949&0.958&0.979&1.03&1.15&1.28&1.79&2.39\\
&&${}^{27}$Al($\alpha$,n)${}^{30}$P&0.787&0.952&0.994&1.0&1.01&1.01&1.22&3.14\\
&&${}^{39}$K(p,$\alpha$)${}^{36}$Ar&1.02&1.01&1.01&0.991&0.941&0.849&1.35&2.64\\
&&${}^{42}$Ca(p,$\alpha$)${}^{39}$K&1.17&1.08&1.02&0.983&0.966&0.961&1.93&2.74\\
&&${}^{44}$Ti(p,$\gamma$)${}^{45}$V&0.864&0.923&0.978&1.02&1.05&1.14&0.69&1.77\\
&&${}^{41}$Sc(p,$\gamma$)${}^{42}$Ti&1.15&1.03&1.0&1.0&1.02&1.08&0.58&1.56\\
&&${}^{33}$S(n,$\alpha$)${}^{30}$Si&0.89&0.967&0.995&1.0&1.01&1.01&1.69&3.11\\
&&${}^{11}$B($\alpha$,n)${}^{14}$N&0.9&0.969&0.997&1.0&1.01&1.0&2.13&3.19\\
&&${}^{45}$V(p,$\gamma$)${}^{46}$Cr&0.902&0.906&0.956&1.04&1.09&1.11&0.93&2.39\\
\hline${}^{47}$Sc&6.12e-08&${}^{47}$Sc(n,$\gamma$)${}^{48}$Sc&0.361&0.531&0.849&1.12&1.25&1.29&1.37&2.6\\
&&${}^{25}$Mg($\alpha$,n)${}^{28}$Si&2.61&1.92&1.24&0.83&0.66&0.619&1.42&2.6\\
&&${}^{47}$Ti(n,p)${}^{47}$Sc&0.408&0.546&0.813&1.24&1.85&2.23&1.38&2.71\\
&&${}^{24}$Mg(n,$\gamma$)${}^{25}$Mg&0.462&0.537&0.811&1.21&1.65&1.91&1.39&2.65\\
&&${}^{30}$Si(p,$\gamma$)${}^{31}$P&2.08&1.77&1.21&0.866&0.745&0.716&2.13&2.65\\
&&${}^{28}$Si(n,$\gamma$)${}^{29}$Si&0.49&0.744&0.944&1.04&1.07&1.08&1.39&2.6\\
&&${}^{27}$Al($\alpha$,p)${}^{30}$Si&0.495&0.631&0.867&1.13&1.35&1.44&1.39&2.65\\
&&${}^{27}$Al(p,$\gamma$)${}^{28}$Si&2.01&1.43&1.07&0.964&0.934&0.926&2.16&2.65\\
&&${}^{20}$Ne($\alpha$,$\gamma$)${}^{24}$Mg&0.503&0.693&0.898&1.1&1.31&1.41&1.41&2.65\\
&&${}^{25}$Mg(p,$\gamma$)${}^{26}$Al&0.51&0.619&0.874&1.1&1.2&1.23&1.41&2.6\\
&&${}^{46}$Sc(n,$\gamma$)${}^{47}$Sc&1.95&1.55&1.15&0.892&0.778&0.748&2.27&2.65\\
&&${}^{25}$Mg(n,$\gamma$)${}^{26}$Mg&0.522&0.671&0.897&1.09&1.2&1.23&1.41&2.6\\
&&${}^{28}$Al(p,$\alpha$)${}^{25}$Mg&1.87&1.37&1.06&0.964&0.934&0.927&2.21&2.6\\
&&${}^{24}$Mg($\alpha$,$\gamma$)${}^{28}$Si&1.8&1.27&1.04&0.981&0.966&0.963&2.18&2.65\\
&&${}^{46}$Sc(p,n)${}^{46}$Ti&0.556&0.692&0.895&1.1&1.29&1.38&1.41&2.65\\
&&${}^{27}$Al(p,$\alpha$)${}^{24}$Mg&0.929&0.93&0.962&1.07&1.36&1.77&2.17&2.65\\
&&${}^{16}$O($\alpha$,$\gamma$)${}^{20}$Ne&1.1&1.07&1.02&0.989&0.912&0.568&1.41&2.65\\
&&${}^{20}$Ne(n,$\gamma$)${}^{21}$Ne&0.576&0.766&0.95&1.03&1.06&1.07&1.41&2.6\\
&&${}^{46}$Ca(n,$\gamma$)${}^{47}$Ca&1.73&1.34&1.07&0.962&0.928&0.92&2.1&2.51\\
&&${}^{45}$Sc(n,$\gamma$)${}^{46}$Sc&1.46&1.37&1.14&0.859&0.654&0.585&1.41&2.65\\
&&${}^{27}$Al(n,$\gamma$)${}^{28}$Al&1.68&1.17&1.02&0.99&0.982&0.98&2.18&2.6\\
&&${}^{26}$Mg($\alpha$,n)${}^{29}$Si&1.22&1.19&1.09&0.891&0.688&0.613&1.42&2.6\\
&&${}^{45}$Sc(p,$\gamma$)${}^{46}$Ti&0.623&0.787&0.945&1.04&1.09&1.11&1.63&2.6\\
&&${}^{29}$Si(n,$\gamma$)${}^{30}$Si&0.632&0.83&0.966&1.02&1.04&1.04&2.06&2.6\\
&&${}^{44}$Ca(p,$\gamma$)${}^{45}$Sc&1.55&1.29&1.08&0.939&0.861&0.835&2.21&2.6\\
&&${}^{45}$Ca(n,$\gamma$)${}^{46}$Ca&1.49&1.31&1.08&0.948&0.898&0.886&2.38&2.6\\
&&${}^{29}$Si(p,$\gamma$)${}^{30}$P&0.672&0.777&0.938&1.04&1.09&1.1&2.12&2.6\\
&&${}^{42}$Ca(p,$\alpha$)${}^{39}$K&1.4&1.08&1.01&0.995&0.991&0.99&2.2&2.65\\
&&${}^{32}$S(n,$\alpha$)${}^{29}$Si&1.37&1.29&1.11&0.916&0.836&0.82&2.15&2.6\\
&&${}^{44}$Ca(n,$\gamma$)${}^{45}$Ca&1.36&1.27&1.08&0.943&0.884&0.869&2.24&2.6\\
&&${}^{26}$Mg(p,$\gamma$)${}^{27}$Al&0.734&0.871&0.983&0.999&0.974&0.96&2.45&2.6\\
&&${}^{41}$Ca(n,$\gamma$)${}^{42}$Ca&1.36&1.15&1.02&0.987&0.976&0.973&2.21&2.6\\
&&${}^{23}$Na($\alpha$,p)${}^{26}$Mg&0.736&0.875&0.979&1.01&1.0&0.997&2.14&2.66\\
&&${}^{46}$Sc(p,$\gamma$)${}^{47}$Ti&0.755&0.948&0.993&1.0&1.01&1.01&2.16&2.6\\
&&${}^{29}$Si($\alpha$,$\gamma$)${}^{33}$S&0.758&0.941&0.992&1.0&1.01&1.01&2.15&2.6\\
&&${}^{32}$S(n,$\gamma$)${}^{33}$S&0.76&0.835&0.945&1.06&1.17&1.22&2.17&2.65\\
&&${}^{46}$Ca(p,$\gamma$)${}^{47}$Sc&1.28&1.13&1.04&0.955&0.862&0.82&2.09&2.6\\
&&${}^{30}$Si(n,$\gamma$)${}^{31}$Si&1.21&1.04&1.0&0.998&0.996&0.996&2.18&2.6\\
&&${}^{38}$Ar($\alpha$,$\gamma$)${}^{42}$Ca&1.2&1.02&1.0&0.999&0.998&0.998&2.19&2.65\\
&&${}^{47}$Ca(p,n)${}^{47}$Sc&0.95&0.96&0.982&1.03&1.11&1.19&2.19&2.6\\
&&${}^{21}$Ne($\alpha$,n)${}^{24}$Mg&1.02&1.01&1.01&0.991&0.946&0.85&2.4&2.56\\
&&${}^{47}$Sc(p,$\gamma$)${}^{48}$Ti&0.859&0.983&0.998&1.0&1.0&1.0&2.17&2.6\\
&&${}^{28}$Si($\alpha$,$\gamma$)${}^{32}$S&0.861&0.962&0.993&1.0&1.01&1.01&2.21&2.6\\
&&${}^{31}$P($\alpha$,p)${}^{34}$S&0.863&0.962&0.994&1.0&1.01&1.01&2.17&2.6\\
&&${}^{31}$P(p,$\alpha$)${}^{28}$Si&1.04&1.03&1.02&0.976&0.91&0.869&2.16&2.6\\
&&${}^{41}$Ca(n,$\alpha$)${}^{38}$Ar&0.964&0.968&0.984&1.02&1.1&1.15&2.2&2.6\\
&&${}^{27}$Al($\alpha$,n)${}^{30}$P&0.878&0.973&0.997&1.0&1.0&1.0&2.17&2.65\\
&&${}^{42}$Ca(n,$\gamma$)${}^{43}$Ca&1.14&1.11&1.04&0.961&0.91&0.895&2.17&2.6\\
&&${}^{27}$Al($\alpha$,$\gamma$)${}^{31}$P&1.13&1.01&1.0&0.999&0.999&0.999&2.17&2.6\\
&&${}^{43}$Ca(n,$\gamma$)${}^{44}$Ca&1.09&1.08&1.03&0.963&0.905&0.885&2.21&2.6\\
&&${}^{47}$Ca(n,$\gamma$)${}^{48}$Ca&0.886&0.957&0.993&1.0&1.01&1.01&2.25&2.56\\
&&${}^{45}$Sc(p,$\alpha$)${}^{42}$Ca&0.887&0.983&0.998&1.0&1.0&1.0&2.17&2.6\\
&&${}^{46}$Sc(n,p)${}^{46}$Ca&1.13&1.1&1.04&0.971&0.936&0.926&2.14&2.6\\
&&${}^{16}$O(${}^{12}$C,p)${}^{27}$Al&0.891&0.979&0.997&1.0&1.0&1.0&2.15&2.6\\
&&${}^{28}$Al(p,n)${}^{28}$Si&1.01&1.01&1.0&0.994&0.964&0.891&2.21&2.6\\
&&${}^{41}$K(p,$\gamma$)${}^{42}$Ca&1.12&1.04&1.01&0.997&0.994&0.993&2.2&2.6\\
&&${}^{21}$Ne(p,$\gamma$)${}^{22}$Na&0.895&0.973&0.997&1.0&1.0&1.0&2.29&2.6\\
&&${}^{41}$K(n,$\gamma$)${}^{42}$K&1.1&1.07&1.02&0.985&0.97&0.966&2.2&2.6\\
&&${}^{30}$P(n,p)${}^{30}$Si&0.985&0.988&0.994&1.01&1.04&1.1&2.21&2.6\\
\hline${}^{48}$V (\iso{Cr}{48})\tablenotemark{a}&9.24e-05&${}^{48}$Cr($\alpha$,p)${}^{51}$Mn&0.395&0.572&0.834&1.2&1.84&3.13&1.83&3.57\\
&&$\alpha$($2\alpha$,$\gamma$)${}^{12}$C&1.31&0.976&0.962&0.972&0.768&0.588&2.3&5.88\\
&&${}^{12}$C($\alpha$,$\gamma$)${}^{16}$O&1.64&1.17&1.03&0.98&0.963&0.958&2.43&4.98\\
&&${}^{44}$Ti($\alpha$,p)${}^{47}$V&1.2&1.16&1.06&0.932&0.766&0.629&1.1&4.33\\
&&${}^{40}$Ca($\alpha$,$\gamma$)${}^{44}$Ti&1.15&1.12&1.05&0.941&0.793&0.664&2.16&2.84\\
&&${}^{13}$N($\alpha$,p)${}^{16}$O&1.34&1.11&1.02&0.986&0.979&0.991&2.42&4.98\\
&&${}^{47}$V(p,$\gamma$)${}^{48}$Cr&1.02&1.01&1.0&0.995&0.976&0.881&2.01&3.9\\
&&${}^{42}$Ca($\alpha$,$\gamma$)${}^{46}$Ti&1.12&1.03&1.0&0.998&0.996&0.996&2.02&4.98\\
&&${}^{16}$O(${}^{12}$C,p)${}^{27}$Al&0.907&0.937&0.977&1.02&1.04&1.05&2.4&5.77\\
\hline${}^{49}$V&6.08e-06&${}^{12}$C($\alpha$,$\gamma$)${}^{16}$O&2.23&1.27&1.05&0.964&0.933&0.925&0.8&4.78\\
&&${}^{49}$Mn(p,$\gamma$)${}^{50}$Fe&0.803&0.855&0.944&1.07&1.34&2.03&0.61&1.23\\
&&${}^{49}$V(p,$\gamma$)${}^{50}$Cr&0.917&0.931&0.971&1.04&1.19&1.89&0.78&4.01\\
&&${}^{49}$Cr(n,p)${}^{49}$V&0.705&0.797&0.933&1.07&1.25&1.49&0.7&3.9\\
&&${}^{13}$N($\alpha$,p)${}^{16}$O&1.47&1.14&1.03&0.98&0.964&0.973&0.75&4.98\\
&&${}^{44}$Ti($\alpha$,p)${}^{47}$V&1.02&0.994&0.991&1.02&1.12&1.44&0.8&4.1\\
&&${}^{42}$Ca($\alpha$,$\gamma$)${}^{46}$Ti&1.3&1.07&1.01&0.995&0.99&0.989&0.74&4.49\\
&&${}^{49}$Cr(p,$\gamma$)${}^{50}$Mn&0.813&0.875&0.958&1.04&1.12&1.17&0.62&1.21\\
&&${}^{48}$Cr(p,$\gamma$)${}^{49}$Mn&1.15&1.12&1.04&0.955&0.87&0.82&0.58&1.03\\
&&${}^{49}$Cr($\alpha$,p)${}^{52}$Mn&0.826&0.96&0.995&1.0&1.01&1.01&0.7&3.73\\
&&${}^{20}$Ne($\alpha$,$\gamma$)${}^{24}$Mg&0.826&0.846&0.955&1.04&1.1&1.12&0.71&5.17\\
&&$\alpha$($2\alpha$,$\gamma$)${}^{12}$C&1.1&0.93&0.95&1.01&0.932&0.833&0.58&6.13\\
&&${}^{48}$Cr($\alpha$,p)${}^{51}$Mn&0.837&0.918&0.979&1.02&1.04&1.06&0.7&3.75\\
&&${}^{23}$Na($\alpha$,p)${}^{26}$Mg&0.856&0.885&0.972&1.02&1.03&1.04&0.71&5.16\\
&&${}^{16}$O(${}^{12}$C,p)${}^{27}$Al&0.861&0.901&0.967&1.02&1.05&1.06&0.71&4.98\\
&&${}^{23}$Na(p,$\gamma$)${}^{24}$Mg&0.864&0.945&0.992&1.0&1.01&1.01&0.71&5.17\\
&&${}^{49}$Cr(n,$\gamma$)${}^{50}$Cr&0.867&0.98&0.998&1.0&1.0&1.0&0.71&3.69\\
&&${}^{16}$O(${}^{12}$C,$\alpha$)${}^{24}$Mg&0.869&0.916&0.979&1.01&1.03&1.03&0.71&4.98\\
&&${}^{27}$Si(n,${}^{12}$C)${}^{16}$O&0.891&0.956&0.993&1.0&1.01&1.01&0.71&4.98\\
&&${}^{26}$Al(n,$\alpha$)${}^{23}$Na&0.899&0.949&0.992&1.0&1.01&1.01&0.72&5.17\\
&&${}^{50}$Mn(n,p)${}^{50}$Cr&0.906&0.985&0.998&1.0&1.0&1.0&0.71&3.72\\
\hline${}^{51}$Cr&1.85e-05&${}^{51}$Mn(p,$\gamma$)${}^{52}$Fe&0.306&0.569&0.848&1.18&1.74&3.13&0.96&3.09\\
&&${}^{48}$Cr($\alpha$,p)${}^{51}$Mn&0.969&0.965&0.982&1.03&1.16&1.69&1.1&1.49\\
&&$\alpha$($2\alpha$,$\gamma$)${}^{12}$C&0.948&0.9&0.955&0.999&0.869&0.724&0.73&5.85\\
&&${}^{27}$Al($\alpha$,n)${}^{30}$P&1.37&1.1&1.01&0.993&0.988&0.986&0.87&1.98\\
&&${}^{44}$Ti($\alpha$,p)${}^{47}$V&1.06&1.02&0.999&1.01&1.07&1.28&0.89&4.19\\
&&${}^{12}$C($\alpha$,$\gamma$)${}^{16}$O&1.28&1.11&1.02&0.971&0.963&0.961&0.85&4.98\\
&&${}^{42}$Ca($\alpha$,$\gamma$)${}^{46}$Ti&1.27&1.07&1.01&0.994&0.988&0.987&0.86&4.33\\
&&${}^{55}$Co(p,$\gamma$)${}^{56}$Ni&0.847&0.927&0.985&1.0&0.969&0.837&0.82&3.23\\
&&${}^{13}$N($\alpha$,p)${}^{16}$O&1.17&1.06&1.02&0.986&0.976&0.988&0.84&5.17\\
&&${}^{33}$S(n,$\alpha$)${}^{30}$Si&1.15&1.05&1.01&0.997&0.995&0.994&0.85&1.93\\
&&${}^{53}$Fe(n,p)${}^{53}$Mn&0.96&0.965&0.983&1.02&1.09&1.14&0.84&3.82\\
&&${}^{53}$Mn(p,$\alpha$)${}^{50}$Cr&0.878&0.97&0.996&1.0&1.0&1.0&0.82&3.71\\
&&${}^{20}$Ne($\alpha$,$\gamma$)${}^{24}$Mg&0.879&0.897&0.964&1.03&1.08&1.1&0.82&5.17\\
&&${}^{51}$Cr(p,$\gamma$)${}^{52}$Mn&0.987&0.989&0.996&1.01&1.03&1.12&0.84&3.81\\
&&${}^{50}$Cr(p,$\gamma$)${}^{51}$Mn&1.08&1.06&1.02&0.981&0.939&0.896&0.64&2.72\\
&&${}^{49}$Mn(p,$\gamma$)${}^{50}$Fe&1.02&1.01&1.0&0.998&1.03&1.12&0.8&1.38\\
&&${}^{45}$Sc(p,$\alpha$)${}^{42}$Ca&1.11&1.07&1.02&0.988&0.976&0.972&0.85&4.33\\
&&${}^{16}$O(${}^{12}$C,p)${}^{27}$Al&0.899&0.932&0.974&1.01&1.03&1.03&0.82&4.98\\
&&${}^{48}$Cr(p,$\gamma$)${}^{49}$Mn&1.02&1.02&1.01&0.987&0.949&0.903&0.69&1.16\\
&&${}^{54}$Fe(p,$\gamma$)${}^{55}$Co&1.0&1.0&1.0&0.998&0.982&0.904&0.82&2.58\\
&&${}^{16}$O(${}^{12}$C,$\alpha$)${}^{24}$Mg&0.906&0.942&0.985&1.01&1.02&1.02&0.82&4.98\\
&&${}^{52}$Fe($\alpha$,p)${}^{55}$Co&0.906&0.971&0.995&1.0&1.01&1.01&0.82&3.71\\
\hline${}^{52}$Mn&0.001&${}^{12}$C($\alpha$,$\gamma$)${}^{16}$O&1.55&1.21&1.04&0.976&0.956&0.951&3.9&4.98\\
&&${}^{13}$N($\alpha$,p)${}^{16}$O&1.39&1.15&1.03&0.979&0.953&0.941&3.9&4.98\\
&&${}^{52}$Fe($\alpha$,p)${}^{55}$Co&0.732&0.871&0.962&1.04&1.15&1.27&2.72&3.63\\
&&${}^{44}$Ti($\alpha$,p)${}^{47}$V&1.09&1.05&1.02&0.98&0.925&0.858&2.56&4.33\\
&&$\alpha$($2\alpha$,$\gamma$)${}^{12}$C&1.11&0.856&0.917&0.993&0.918&0.863&2.27&5.98\\
&&${}^{53}$Fe(n,p)${}^{53}$Mn&0.937&0.947&0.977&1.03&1.1&1.15&3.38&3.82\\
&&${}^{16}$O(${}^{12}$C,p)${}^{27}$Al&0.886&0.925&0.972&1.02&1.05&1.06&3.85&5.15\\
&&${}^{16}$O(${}^{12}$C,$\alpha$)${}^{24}$Mg&0.897&0.935&0.981&1.01&1.03&1.03&3.85&4.98\\
&&${}^{53}$Mn(p,$\alpha$)${}^{50}$Cr&0.899&0.981&0.998&1.0&1.0&1.0&3.28&3.71\\
\hline${}^{53}$Mn&0.000107&${}^{53}$Fe(n,p)${}^{53}$Mn&0.546&0.707&0.902&1.1&1.3&1.42&2.93&3.72\\
&&${}^{12}$C($\alpha$,$\gamma$)${}^{16}$O&1.52&1.2&1.04&0.971&0.959&0.956&3.25&4.78\\
&&${}^{13}$N($\alpha$,p)${}^{16}$O&1.36&1.14&1.03&0.975&0.943&0.931&3.21&4.78\\
&&${}^{52}$Fe(p,$\gamma$)${}^{53}$Co&1.23&1.09&1.02&0.987&0.969&0.961&0.74&1.01\\
&&${}^{53}$Mn(p,$\gamma$)${}^{54}$Fe&0.956&0.976&0.992&1.01&1.04&1.19&3.15&3.81\\
&&${}^{42}$Ca($\alpha$,$\gamma$)${}^{46}$Ti&1.15&1.05&1.01&0.995&0.991&0.99&3.56&4.49\\
&&${}^{52}$Fe($\alpha$,p)${}^{55}$Co&0.87&0.949&0.988&1.01&1.03&1.05&2.8&3.63\\
&&${}^{52}$Fe(n,p)${}^{52}$Mn&1.08&1.06&1.02&0.975&0.916&0.873&3.06&3.56\\
&&${}^{16}$O(${}^{12}$C,p)${}^{27}$Al&0.873&0.909&0.965&1.03&1.06&1.07&2.99&4.78\\
&&${}^{16}$O(${}^{12}$C,$\alpha$)${}^{24}$Mg&0.881&0.921&0.977&1.01&1.03&1.03&2.98&4.98\\
&&${}^{27}$Si(n,${}^{12}$C)${}^{16}$O&0.9&0.954&0.993&1.0&1.01&1.01&2.99&4.78\\
&&${}^{52}$Mn(p,$\gamma$)${}^{53}$Fe&1.0&1.0&1.0&0.997&0.983&0.906&3.06&3.62\\
&&${}^{53}$Fe(n,$\gamma$)${}^{54}$Fe&0.907&0.989&0.999&1.0&1.0&1.0&3.05&3.56\\
\hline${}^{55}$Fe&0.000495&${}^{55}$Co(p,$\gamma$)${}^{56}$Ni&0.605&0.761&0.921&1.09&1.32&1.8&2.21&3.13\\
&&${}^{12}$C($\alpha$,$\gamma$)${}^{16}$O&1.53&1.23&1.05&0.96&0.939&0.932&3.81&5.17\\
&&${}^{13}$N($\alpha$,p)${}^{16}$O&1.39&1.15&1.04&0.97&0.929&0.907&3.9&5.6\\
&&${}^{16}$O(${}^{12}$C,p)${}^{27}$Al&0.858&0.893&0.957&1.03&1.06&1.07&3.57&5.15\\
&&${}^{16}$O(${}^{12}$C,$\alpha$)${}^{24}$Mg&0.864&0.907&0.972&1.02&1.03&1.04&3.58&4.49\\
&&${}^{42}$Ca($\alpha$,$\gamma$)${}^{46}$Ti&1.15&1.05&1.01&0.995&0.99&0.989&3.56&4.49\\
&&${}^{12}$C(${}^{12}$C,$\alpha$)${}^{20}$Ne&1.14&1.09&1.02&0.984&0.967&0.963&3.19&5.17\\
&&${}^{12}$C(${}^{12}$C,p)${}^{23}$Na&1.14&1.07&1.02&0.989&0.978&0.975&3.36&5.17\\
&&${}^{27}$Si(n,${}^{12}$C)${}^{16}$O&0.886&0.946&0.992&1.0&1.01&1.01&3.65&4.49\\
&&${}^{23}$Na($\alpha$,p)${}^{26}$Mg&0.887&0.894&0.965&1.02&1.04&1.04&3.54&5.16\\
&&${}^{54}$Fe(p,$\gamma$)${}^{55}$Co&1.0&1.0&1.0&1.0&1.02&1.12&2.37&2.83\\
\hline${}^{56}$Co (\iso{Ni}{56})\tablenotemark{a}&0.0875&$\alpha$($2\alpha$,$\gamma$)${}^{12}$C&1.24&1.14&1.05&0.944&0.847&0.803&3.22&5.77\\
&&${}^{12}$C($\alpha$,$\gamma$)${}^{16}$O&1.11&1.04&1.01&0.996&0.993&0.992&3.99&4.98\\
\hline${}^{57}$Co (\iso{Ni}{57})\tablenotemark{a}&0.00256&${}^{57}$Ni(n,p)${}^{57}$Co&0.727&0.809&0.934&1.07&1.24&1.38&2.04&3.48\\
&&$\alpha$($2\alpha$,$\gamma$)${}^{12}$C&1.1&1.1&1.02&0.978&0.915&0.87&2.55&5.55\\
&&${}^{56}$Ni(p,$\gamma$)${}^{57}$Cu&1.14&1.03&1.01&0.996&0.991&0.989&0.46&0.64\\
&&${}^{27}$Al($\alpha$,n)${}^{30}$P&1.11&1.03&1.0&0.998&0.996&0.995&2.64&3.22\\
&&${}^{57}$Ni(n,$\gamma$)${}^{58}$Ni&0.907&0.986&0.998&1.0&1.0&1.0&2.93&3.31\\
\hline${}^{59}$Fe&3.92e-05&${}^{59}$Fe(n,$\gamma$)${}^{60}$Fe&0.687&0.843&0.967&1.02&1.04&1.04&1.32&2.34\\
&&${}^{58}$Fe(n,$\gamma$)${}^{59}$Fe&1.31&1.15&1.04&0.964&0.903&0.871&1.32&2.34\\
&&${}^{24}$Mg(n,$\gamma$)${}^{25}$Mg&0.825&0.859&0.946&1.05&1.12&1.14&1.32&2.34\\
&&${}^{25}$Mg($\alpha$,n)${}^{28}$Si&1.21&1.15&1.05&0.966&0.93&0.921&1.34&2.34\\
&&${}^{20}$Ne($\alpha$,$\gamma$)${}^{24}$Mg&0.832&0.879&0.962&1.03&1.07&1.08&1.32&2.34\\
&&${}^{20}$Ne(n,$\gamma$)${}^{21}$Ne&0.844&0.918&0.983&1.01&1.02&1.02&1.32&2.34\\
&&${}^{16}$O($\alpha$,$\gamma$)${}^{20}$Ne&1.07&1.06&1.03&0.965&0.891&0.849&1.32&2.34\\
&&${}^{25}$Mg(n,$\gamma$)${}^{26}$Mg&0.853&0.905&0.974&1.02&1.04&1.05&1.32&2.34\\
&&${}^{59}$Fe(p,n)${}^{59}$Co&0.858&0.96&0.994&1.0&1.01&1.01&1.32&2.34\\
&&${}^{25}$Mg(p,$\gamma$)${}^{26}$Al&0.871&0.918&0.98&1.01&1.03&1.03&1.32&2.34\\
&&${}^{28}$Al(p,$\alpha$)${}^{25}$Mg&1.13&1.06&1.01&0.994&0.989&0.988&1.34&2.38\\
&&${}^{28}$Si(n,$\gamma$)${}^{29}$Si&0.888&0.962&0.993&1.0&1.01&1.01&1.32&2.34\\
&&${}^{26}$Mg($\alpha$,n)${}^{29}$Si&1.12&1.08&1.03&0.974&0.937&0.925&1.34&2.34\\
&&${}^{27}$Al($\alpha$,p)${}^{30}$Si&0.907&0.951&0.988&1.01&1.02&1.02&1.32&2.34\\
&&${}^{58}$Fe(p,$\gamma$)${}^{59}$Co&0.908&0.97&0.995&1.0&1.01&1.01&1.32&2.34\\
\hline${}^{59}$Ni&0.000162&${}^{59}$Cu(p,$\gamma$)${}^{60}$Zn&0.274&0.464&0.775&1.29&2.12&2.77&1.37&2.39\\
&&${}^{57}$Cu(p,$\gamma$)${}^{58}$Zn&1.68&1.17&1.02&0.989&0.982&0.974&0.78&1.07\\
&&${}^{59}$Cu(p,$\alpha$)${}^{56}$Ni&1.41&1.25&1.06&0.963&0.93&0.963&1.25&2.39\\
&&${}^{57}$Ni(n,p)${}^{57}$Co&0.856&0.894&0.955&1.06&1.21&1.39&1.19&2.96\\
&&$\alpha$($2\alpha$,$\gamma$)${}^{12}$C&0.739&0.893&0.964&1.01&0.924&0.847&1.08&5.72\\
&&${}^{20}$Ne($\alpha$,$\gamma$)${}^{24}$Mg&0.937&0.948&0.98&1.03&1.13&1.31&1.73&2.04\\
&&${}^{57}$Ni(p,$\gamma$)${}^{58}$Cu&1.01&1.01&1.0&0.989&0.926&0.788&1.21&1.6\\
&&${}^{27}$Al($\alpha$,n)${}^{30}$P&1.25&1.06&1.01&0.995&0.994&1.0&1.19&1.54\\
&&${}^{48}$Cr($\alpha$,p)${}^{51}$Mn&0.974&0.979&0.993&1.01&1.08&1.22&1.18&1.51\\
&&${}^{58}$Ni(p,$\gamma$)${}^{59}$Cu&1.02&1.01&1.0&0.997&0.961&0.849&1.29&2.4\\
&&${}^{30}$Si(p,$\gamma$)${}^{31}$P&0.981&0.978&0.992&1.01&1.08&1.14&1.37&2.6\\
&&${}^{33}$S(n,$\alpha$)${}^{30}$Si&1.14&1.04&1.0&1.0&0.999&1.0&1.27&1.96\\
&&${}^{58}$Cu(p,$\gamma$)${}^{59}$Zn&1.0&1.0&1.0&1.01&0.972&0.882&1.24&1.63\\
&&${}^{26}$Al(n,$\alpha$)${}^{23}$Na&1.13&1.03&1.0&1.0&0.996&0.994&1.18&1.92\\
&&${}^{58}$Ni($\alpha$,$\gamma$)${}^{62}$Zn&0.905&0.973&0.993&0.997&1.01&1.01&1.27&2.65\\
\enddata
\tablenotetext{a}{Sensitivity of long-lived parent isotope is the same}
\end{deluxetable*}
\bibliography{ccsnRxnSens}



\end{document}